\documentclass[frenchb,twoside,fleqn]{mn2e}

\input psfig.sty
\usepackage{times}

\def\R#1{{\mathrm{#1}}}		
\def\Eq#1{{Eq.~(\ref{e:#1})}}	
\def\Ep#1{{~(\ref{e:#1})}}	
\def\Eqs#1#2{{Eqs.~(\ref{e:#1})-(\ref{e:#2})}}
\def\EQN#1{\label{e:#1}}        
\def\Fig#1{{Fig.~(\ref{f:#1})}}

\def\be{\begin{equation}}
\def\ee{\end{equation}}
\def\ba{\begin{eqnarray}}
\def\ea{\end{eqnarray}}

\def\M#1{{\mathbf{#1}}}	
\def\T#1{{{#1}^{\bot}}}		
\def\d#1{{\R{d}{#1}}}		
\def\mdot{\!\cdot\!}		

\newcommand{\nicefrac}[2]{\leavevmode\kern.1em
            \raise.5ex\hbox{\the\scriptfont0 #1}\kern-.1em
      /\kern-.15em\lower.25ex\hbox{\the\scriptfont0 #2}}

\title[ 
On the kinematic deconvolution    
of the local   neighbourhood luminosity function  ]
{ 
On the kinematic deconvolution    \\  
of the local   neighbourhood luminosity function \\ }

\author[   C.~Pichon, A.~Siebert, \& O.Bienaym\'e   ]
{
  C.~Pichon${}^{1,2}$, A.~Siebert${}^{1}$ \& O.~Bienaym\'e${}^{1}$ \\ 
${}^1$ Observatoire de Strasbourg, 11 rue de l'Universit\'e,
 67000 Strasbourg, France. \\
${}^2$ Institut d'Astrophysique de Paris, 98 bis boulevard
       d'Arago, 75014 Paris, France. \\
 }

\date{\today}

\begin{document}

\maketitle

\label{firstpage}

\maketitle
\markboth{ On the kinematic deconvolution of the local luminosity function    }{
  }


\begin{abstract}
A  method for  inverting  the statistical  star  counts equation,  including
proper  motions, is  presented; in  order to  break the  degeneracy  in that
equation  it  uses  the  supplementary  constraints  required  by  dynamical
consistency.   The inversion  gives access  to both  the kinematics  and the
luminosity  function of  each population  in three  r\'egimes:  the singular
ellipsoid, the constant ratio  Schwarzschild ellipsoid plane parallel models
and the epicyclic  model.  This more realistic model  is taylored to account
for local neighbourhood density  and velocity distribution.

  The first model
is fully  investigated both analytically  and via means of  a non-parametric
inversion  technique, while the  second model  is shown  to be  formally its
equivalent.  The effect of noise and incompleteness in apparent magnitude is
investigated.  The  third model is  investigated via a  5D+2D non-parametric
inversion technique  where positivity of the  underlying luminosity function
is explicitely accounted for.

It is argued that its future application to data such as the Tycho catalogue
(and in  the upcoming  satellite GAIA) could  lead -- provided  the vertical
potential,  and/or the  asymmetric  drift or  $w_\odot$  are known  -- to  a
non-parametric determination of  the local neighbourhood luminosity function
without any reference to stellar evolution tracks.  It should also yield the
proportion of stars for each  kinematic component and a kinematic diagnostic
to split the thin disk from the thick disk or the halo.
\end{abstract}



%

\section{Introduction}

Most of  our knowledge of the global  structure of the Galaxy  relies on the
comparison  of  magnitude  and  colour  star counts  in  different  Galactic
directions. Star  counts alone do not allow  us to solve the  dilemma that a
star of  a given  apparent magnitude can  be either intrinsically  faint and
close by, or bright and distant. This problem may be addressed statistically
by using the century-old equation of stellar statistics (von Seeliger 1898):

\begin{equation}    
A_{\lambda}(m,\ell,b)= \int_0^\infty \Phi_{\lambda}(M)  \, 
\rho(r,\ell,b)\, r^2 \, \d r
\, ,  \EQN{eqstat}
\end{equation}

where $A_{\lambda}(m,\ell,b) \, \d  m \,  \d \ell \, \d  (\sin b)$ is the
number of stars which have an apparent magnitude in the range $[ m,m+ \d m]$\,,
$\Phi_{\lambda}(M)$ is the luminosity function, which depends on the
intrinsic magnitude, $M$, and the colour band $\lambda$, while $\rho(r,\ell,b)$
is the density at radius $r$ (within $\d r$) along the line of sight in the
direction given by the Galactic longitudes and latitudes $(\ell,b)$ (within 
the solid angle $\d \ell \cos(b) \d b$).

This equation  cannot be  solved or inverted  (i.e. by determining  both the
stellar  LF and  the density  law) except  for a  few simplified  cases. For
instance,  with a  ``homogeneous''  stellar sample  for  which the  absolute
magnitudes of stars or more  precisely their luminosity functions are known,
the  density law  along the  line of  sight can  be recovered.   A classical
numerical technique  (Mihalas \& Binney 1981)  has been proposed  -- the Bok
diagram  (1937) --  while more  rigorous treatments  are required  for small
samples to stabilize the inversion so as to produce smooth solutions (Binney
\& Merrifield  1998).  The  converse situation is  the determination  of the
luminosity function assuming a known  density law (see, for instance, recent
studies of  the faint end of  the disk or  halo main sequence based  on deep
star counts (Reid et al. 1996, Gould et al. 1998).

A  simple approach,  developed largely  in  the eighties,  was to  integrate
\Eq{eqstat}   assuming  some  prior   information  concerning   the  stellar
populations (see, for  instance, Pritchet 1983, Bahcall et  al., 1983, Buser
1985, Robin \&  Cr\'ez\'e 1986). A frequent assumption  is, for instance, to
assume that  the halo stars  have the same  luminosity function as  some low
metallicity  globular clusters.   Another  approach consists  in building  a
stellar luminosity function from  stellar evolution tracks and isochrones of
various ages.   This has been used  to put constraints on  the Galactic disk
star formation rate (Haywood et al 1997ab).

Stronger a  priori constraints  may also be  derived by  requiring dynamical
consistency,  since the  vertical  kinematics  of stars  is  related to  the
flattening of stellar disks or spheroidal components.

Since  star counts alone,  $A_{\lambda}(m, \ell,b)$,  are not  sufficient to
constrain uniquely  Galactic stellar population models, it  is expected that
two (or more) distinct models  will reproduce the same apparent star counts.
However, this  is not a  real worry,   since  it is  likely that  adding some
relevant  extra  a  priori  information  must help  to  lift  partially  the
degeneracy of the models.

In this paper, it is shown that the degeneracy is  lifted altogether when we
consider, in addition to the  star counts in  apparent magnitude, the proper
motions, $\mu_{\ell}$  and $\mu_{b}$.  For  a relatively general dynamically
consistent  model  (stationary,  axisymmetric  and   fixed kinematic  radial
gradients), the  statistical equation counts may  be formally inverted, giving
access to {\sl both} the vertical density law of each stellar population  
{\sl and} their  luminosity   functions.        This is   developed    in
section~\ref{s:deriv} where we show how the vertical motions are related to
the thickness of stellar  components.  The remaining degeneracy occurs  only
for   a quadratic vertical   potential.   Otherwise ---   when the  vertical
component   of the potential   is known  ---  the  departures from quadratic
behaviour   define  a  characteristic scale   that  allows  us to  transform
statistically    the  magnitudes into   distances  and   proper motions into
velocities.  Similarly, the  asymmetric  drift and/or the  vertical velocity
component of the Sun provide a natural scale in  energy, leading to the same
inversion procedure.

For ideal star counts (infinitely deep and for an infinite number of stars),
the  inversion gives  exactly  the  proportion of  stars  in each  kinematic
component, providing  a direct  diagnostic to split  the thin disk  from the
thick disk or  the halo, and its luminosity  function $\Phi_{\lambda}(M)$ is
recovered  for   each  kinematic  stellar  component.   This   is  a  direct
consequence of  the supplementary constraints introduced  by the requirement
for dynamical consistency.

Section~\ref{s:deriv}  presents the  generalized stellar  statistic equation
which accounts  for proper motions,  and demonstrates the uniqueness  of the
inversion  for two families  of plane  parallel distribution  functions: the
singular  velocity ellipsoid (subsection~\ref{s:toy})  and a  constant ratio
velocity  ellipsoid  (subsection~\ref{s:sch}) while  {subsection~\ref{s:epi}
presents a basic description  of the epicyclic model}. Section~\ref{s:simul}
illustrates  the inversion  procedure  on a  fictitious  superposition of  4
kinematically  decoupled populations  with distinct  main  sequence turn-off
magnitudes  {for the  constant ratio  velocity ellipsoid  and  the epicyclic
models.  The  next section discusses  the effects of truncation  in apparent
magnitude (i.e.  completude of the catalog)  in the recovered LF  as well as
noise  in   the  measurements.   Finally,  the  last   section  discuss  the
applicability  of  the method  to  the  Tycho-2  catalogue and  to  external
clusters, and concludes.}

\onecolumn

\section{Derivation}
\label{s:deriv}

The number of stars, $\d N$, which have an apparent luminosity in the range $[
L,L+ \d L[$\, in the  solid  angle defined by   the Galactic longitudes  and
latitudes $(\ell,b)$ (within $ \d \ell \d {(\sin  b)}$), with proper motions
$\mu_{\ell}$ and $\mu_{b}$ (within $\d{}  \mu_{b}$ and $\d{} \mu_{\ell}$) is
given by

\begin{equation}
\d N  \equiv A_{\lambda}(L,
\mu_l,\mu_b;l,b)\, \d \mu_{\ell} \d \mu_{b} \d {\ell} \cos{b} \d b \d 
L  = \left\{
  {\int \!\!  {\int  {
\Phi_{\lambda}^{\star}\left[ L_0 ,\beta \right]}}\, \left(\int \!\! 
f_\beta(\M{r},\M{u})\d u_r \right) r^4 
  \d r\d\beta
} \right\} \, \d \mu_{\ell} \d \mu_{b} \d {\ell} \cos{b} \d b \d L  
\, ,\EQN{N0}
\end{equation}

where  we have  introduced the  luminosity  function per  unit bandwidth,  $
\Phi^{\star}_\lambda\left[ L_0 ,\beta \right] $, which is here taken to be a
function of the absolute luminosity,  $L_{0}$, and of a continuous kinematic
index,  $\beta$.  The variables  $\M{r},\M{u}$ are  the vector  position and
velocity coordinates ($u_r,u_\ell,u_b$) in phase space relative to the Local
Standard  of Rest,  while  $\M{R}$ and  $\M{V}$  are those  relative to  the
Galactic centre.  The relationship between $A_{\lambda}(L, \mu_l,\mu_b;l,b)$
and  $\Phi_\lambda^{\star}\left[  L_0 ,\beta  \right]  $  involves a  double
summation over $\beta  $, and distance, $r$, along the  line of sight.  Here
$f_\beta(\M{r},\M{u})$ represents the  $\beta$ component of the distribution
function of the assumed stationary axisymmetric equilibrium, {\it i.e.}

\begin{equation}
f(\M{r},\M{u}) = \int_{0}^{\infty} f_\beta(\M{r},\M{u}) \, \d \beta
\EQN{decompdf}
\end{equation}

where $f$ is decomposed over  the basis of isothermal solutions $f_\beta$ of
the   Boltzmann   equation  for   the   assumed   known  potential   $\psi$.
\Eq{decompdf} corresponds to a  decomposition over isothermal populations of
different   kinematic  temperatures,   $\sigma^{2}  =1/\beta$.   Apart  from
restriction, the shape of the distribution $f(E_z)$ could be anything.  Note
that \Eq{N0} is a direct generalization of \Eq{eqstat} since

\[
\rho_\beta(r,\ell,b, \mu_\ell,\mu_b) \equiv
r^{2} \int \!\! 
f_\beta(\M{r},\M{u})\d u_r  \,
\] 

is  by definition  the density  of stars  (belonging to  population $\beta$)
which are  at position  $\M{r} \equiv  (r,\ell,b)$ within $\d  r \d  \ell \d
{(\sin  b)}$,  with proper  motion  $\mu_{b}$  (within  $\d{} \mu_{b}$)  and
$\mu_{\ell}  $(within $\d{}  \mu_{\ell}$).  The  extra summation  on $\beta$
which  arises in  \Eq{N0} accounts  for  the fact  that stars  in the  local
neighbourhood come  from a superposition of  different kinematic populations
which,   as   is   shown    later,   can   be   disentangled.    Note   that
$\Phi_{\lambda}^{\star}$  is  defined  here  per unit  absolute  luminosity,
$L_{0}$, and therefore

\[
\Phi_{\lambda}(M(L_{0}))= \frac{2 \log(10)}{5}{L_{0}}\int 
\Phi_{\lambda}^{\star}[L_{0},\beta]\d \beta \quad{\rm where} \quad 
M(L_{0})= -\frac{5}{2}\frac{1}{\log 10} \log\left( 
\frac{L_{0}}{L_{\odot}}\right)+M_\odot \, .
\]

Since  there is  no convolution  on  $\lambda$ (which  is mute)  it will  be
omitted from now on in the derivation.  In section~\ref{s:simul} B-V colours
are reintroduced to  demonstrate the inversion for a  fictitious HR diagram.
We shall  also drop the $\star$ superscript  but will keep in  mind that the
luminosity function is  expressed as a function of  the absolute luminosity,
$L_{0}$.

This paper is concerned with the  inversion of \Eq{N0}.  We proceed in three
steps; first  a simplistic Ansatz  for the distribution function  is assumed
(corresponding  to  a stratification  in  height  of  uniform disks  with  a
pin-like  singular velocity  ellipsoid) leading  to  a proof  that, in  this
context,  \Eq{N0} has  a  well-defined  unique solution  which  can be  made
formally explicit.  A more realistic  model is then presented accounting for
the measured anisotropy of the velocity ellipsoid.  It is shown that, in the
direction of the Galactic centre, and if the velocity dispersions ratios are
constant for  all populations, this  model is formally  invertible following
the  same  route.   Away  from  the Galactic  centre  direction,  the  Sun's
velocities  are also accounted  for to  recover statistically  distances via
another  inversion  procedure related  to  secular  parallaxes. Finally,  we
illustrate  the inversion  on a  fully 7  dimensional epicyclic  model.  The
detailed investigation  of this model  is postponed to a  companion paper,
(Siebert, Pichon, \& Bienaym\'e in preparation).

\subsection{ A toy model: parallel sheet model with singular velocity ellipsoid}
\label{s:toy}

Let  us assume here   a sheet-like  model  for the distribution  function of
kinematic temperature $\beta$:
\begin{equation}
f_\beta(\M{r},\M{u})=\sqrt{\frac{\beta}{2\pi}} \exp \left(  {-\beta
E_z }   \right)\delta(v_R)\delta(v_\phi) \, ,\EQN{ansatzPP}
\end{equation}
which corresponds  to a  stratification in height  with a  pin-like singular
velocity ellipsoid  which is aligned with  the rotation axis  of the Galaxy.
Calling  $\mu_{b}  \equiv   u_b/r$,  the  energy  reads  in   terms  of  the
heliocentric coordinates
\begin{equation}
E_z =\frac{v_z^2}{2}+\psi_{z}(z) =
 { {r^2 \mu_b }^2 \over { 2 \cos^2 (b)}}+ \alpha \, {\sin^2 (b)}
r^2 +\chi \, (r \sin(b)) \,, \EQN{Ez}
\end{equation}
where the  harmonic component of the  $z$ potential ($\alpha  z^2$) was made
explicit while leaving unspecified the non harmonic-residual, $\chi$.

Putting \Eq{ansatzPP} into \Eq{N0} leads to
\begin{equation}
A[b,\mu_b   ,L]=   {\int  \!\!       {\int    {
\sqrt{\frac{\beta}{2\pi}}\frac{\Phi\left[ L  r^2  ,\beta  
\right]}{\cos(b)}}}\exp \left( {-\beta
\alpha r^2\sin^2 (b)-\beta r^2{{\mu_b ^2}  \over {2 \cos^2 (b)}}-\beta \chi( r
\sin(b))} \right)r^3 \d r\d\beta }   \,, \EQN{N1}
\end{equation}
given  the relationship  $L_{0}=  L r^{2}$  relating  apparent and  absolute
luminosities.  Introducing $\zeta= L^{1/2} r$ ,
\begin{equation}
    x = \alpha \frac{\sin^2 (b)}{L}+ {{\mu_b ^2} \over { 2 L \cos^2 (b)}} \, ,
\quad {\rm and} \quad y= \frac{\sin(b)}{L^{1/2}} \, . \EQN{defx}
\end{equation}
\Eq{N1} then reads
\begin{equation}
L^2 \cos(b) A[b,\mu_b   ,L]=     {\int    \!\!   {\int
\sqrt{\frac{\beta}{2\pi}} {\Phi\left[ \zeta^2,\beta \right]}}\exp \left[ {-\beta \zeta^2
x  -\beta \chi( \zeta  y)} \right] \zeta^3  \d \zeta  \d\beta  } \, .
\EQN{N22}
\end{equation}

\subsubsection{harmonic degeneracy}

Suppose for now that the $z$-potential is purely harmonic, so that $\chi$ is
identically null.  Calling  $ s= \beta \zeta^{2} $,  the inner integral over
$\zeta$ in \Eq{N22} can be rewritten as an integral over $\beta$ and $s$
\begin{equation}
{\int \!\!  {\int { \sqrt{\frac{\beta}{2\pi}} \Phi\left[ \zeta^2,\beta \right]}}\exp
\left[ {-\beta \zeta^2 x } \right]\zeta^3 \d \zeta \d\beta }=\frac{1}{2 \sqrt{2\pi}} \int
\!\! \left( \int  \Phi\left[ s/\beta,\beta \right] \beta^{-3/2}\d \beta \right)
\exp \left[ - s x \right] s \d s \, . \EQN{phi0}
\end{equation}
\Eq{phi0} shows  that for a purely harmonic  potential the {\sl  mixture} of
populations (integrated  over $\beta$)  is  recovered  from $A[b,\mu_b  ,L]$
which is effectively a  function of $x$ only  (given by \Eq{defx}).  In this
instance, the inversion does   not  allow us  to disentangle   the different
kinematic populations.  In physical terms  there is a degeneracy between the
distance,  luminosity and proper  motion.   In contrast,  when the  data set
extends far enough to  probe the an-harmonic part  of the potential, we  now
demonstrate   that \Eq{N22} has  formally  a  unique  exact solution, before
exploring non-parametric means of inverting it in a more general framework.

\subsubsection{Uniqueness ?}
\label{s:unique}
Let  us assume
  that not too far from the Galactic plane, $\chi(z)$ is well approximated by
$\chi(z) = \gamma z^\nu$, so that \Eq{N22} becomes
\begin{equation}
L^2 \cos(b) A[b,\mu_b ,L]={\int \!\! {\int \sqrt{\frac{\beta}{2\pi}}
{\Phi\left[  \zeta^2,\beta \right]}}\exp \left( -\beta  \zeta^2 x  -\beta \gamma \zeta^\nu
y^\nu \right)\zeta^3 \d \zeta \d\beta }  \, . \EQN{N2}
\end{equation}
Calling
\begin{equation}
\Phi_1\left[ U,B \right]=\frac{1}{\sqrt{2\pi}}\Phi\left[ \exp(2U) ,\exp  (B)
\right] \exp(4   U+3/2    B-c  ([2+\nu] U+2B)),   \,\,\,\quad   K_0\left(  z
\right)=\exp (c z-\exp(z)) \,, \EQN{defK}
\end{equation}
\begin{equation}
{\rm and } \quad A_1\left[ X,Y  \right]=L^2
\cos b A[b,\mu_b ,L] \exp \left[c(X+Y)\right] \, ,\EQN{defA1}
\end{equation}
where
\begin{equation}
B=\log     (\beta  ),\quad   Z=\log(\zeta),\quad  X=\log(x)=\log\left[   {\alpha
\frac{\sin^2  (b)}{L}+{{\mu_b^2} \over {2 L  \cos^2 (b)}}} \right], \quad Y=
\log\left|     \frac{   {\gamma    \sin^\nu (b)}}{L^{\nu/2}}  \right|,
\EQN{defX}
\end{equation}
\Eq{N2} becomes
\begin{equation}
A_1[b,\mu_b   ,L]=A_1[X,Y]=\int   \!\! \int   
 {\Phi_1\left[ {Z,B}  \right]}K_0\left(
{B+2Z+X} \right)K_0\left( {B+\nu Z+Y} \right) \d Z\d B \, . \EQN{N4}
\end{equation}
The positive scalar $c$ is left to our discretion and can be chosen so as to
yield a narrow  kernel, $K_0$ (in practice c  should be close  to one).  Since
$r$ runs from zero to infinity and so does $\beta$, the integration over $B$
and  $Z$ will run  from $-\infty$ to  $\infty$.  Similarly $X$  and $Y$ span
$]-\infty,\infty[$ as $b$ goes from zero to $\pi/2$.  Let
\begin{equation}
w=-(B+\nu Z),\quad  \varpi=-(B+2Z), 
\end{equation}
\Eq{N4} reads
\begin{equation}
A_1[X,Y]=|\nu-2|^{-1}\int  \!\!{\int  {\Phi_1\left[  {\varpi,w}   \right]}}K_0\left(  
{X-\varpi}
\right)K_0\left( {Y-w} \right)\d \varpi\d  w \,. \EQN{convol} 
\end{equation}
The {\sl unique} solution of \Eq{convol} reads formally
\begin{equation}
\Phi_1\left[    {\varpi,w}          \right]=|\nu-2| \, {\rm    FT}^{-1}\left(   {{{\hat
A_1[k_\varpi,k_w]}       \over     {\hat       K_0(k_\varpi)\hat      K_0(k_w)}}}
\right) \, ,\EQN{deconvol}
\end{equation}
where
\[\hat A_1[k_\varpi,k_w]=\int\!\!\!\int {\exp \left[ {+i\left( 
{k_w X+k_\varpi Y} \right)}  \right]}A_1[X,Y]\d X\d Y  \quad {\rm and} \quad
\hat K_0[k]=\int  {\exp \left[ {+i\left( {k  X}  \right)} \right]K_0\left( X
\right)\d X} \,,
\]
while
\[FT^{-1}\left( {f(k_{x},k_{y})} \right)={1 \over {4\pi^2 }}\int\!\! \int
 {\exp \left[ {-i {k_x \varpi}  -i {k_y w}} \right]f\left( k_x,k_y \right)\d
k_x\d k_y} \, .
\]

Both Fourier  transforms are well-defined  given the span  of $\varpi,w$ and
$X,Y$.   Approximating  both  $K_0$  and   $A_1$  by  a  Gaussian  of  width
respectively $1/\ell_K$  and $1/\ell_N^2$ \Eq{deconvol}  shows that $\Phi_1$
will be a Gaussian of width $1/(\ell_N-\ell_K)^2$.  

This  procedure is  therefore  a {\sl  true}  deconvolution: the  luminosity
function  $\Phi_{\lambda}(L,\beta)$ is  effectively  recovered at  arbitrary
resolution (in effect  fixed by the signal to noise ratio  of the data).  In
practice, \Eq{deconvol}  is impractical  for noisy finite  data sets,  so we
shall  investigate  non-parametric  regularised  solutions  to  \Eq{N22}  in
section~\ref{s:NP}.

\begin{figure*}\unitlength=1cm
\begin{picture}(14,16)
\put(-3.5,8){\centerline{\psfig{width=9cm,file=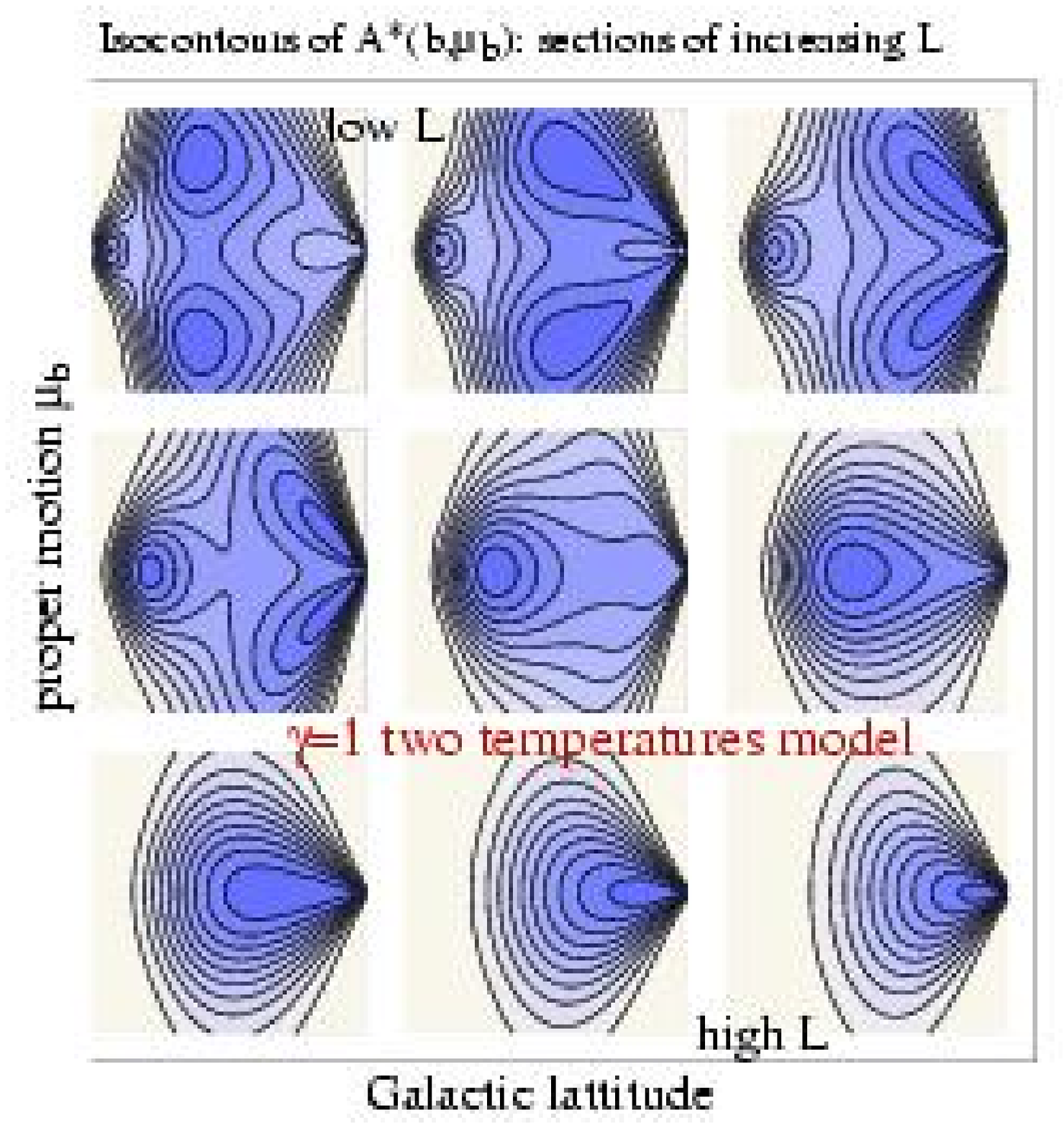}}}
\put(4.5,8){\centerline{\psfig{width=9cm,file=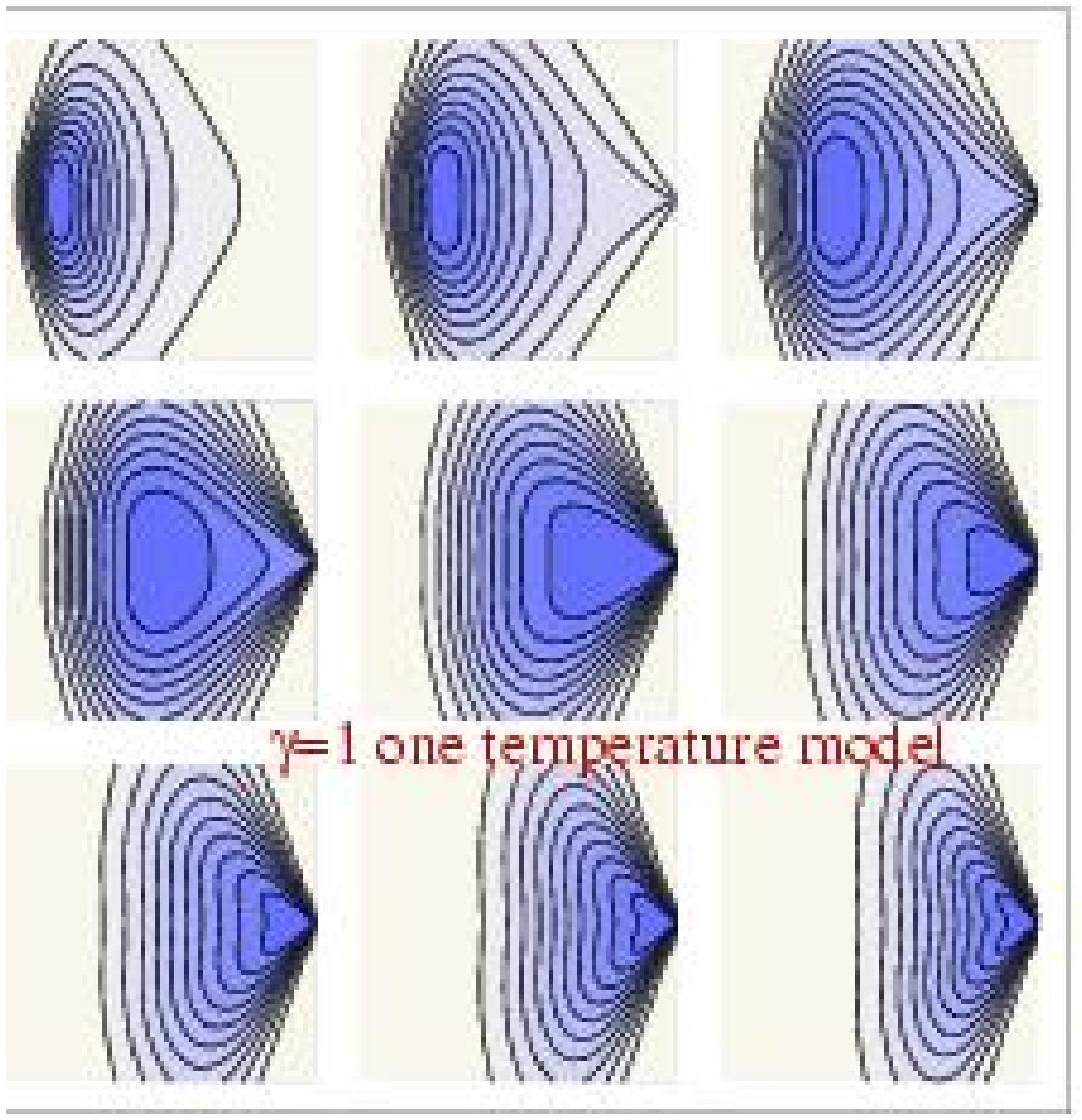}}}
\put(-3.5,-0){\centerline{\psfig{width=9cm,file=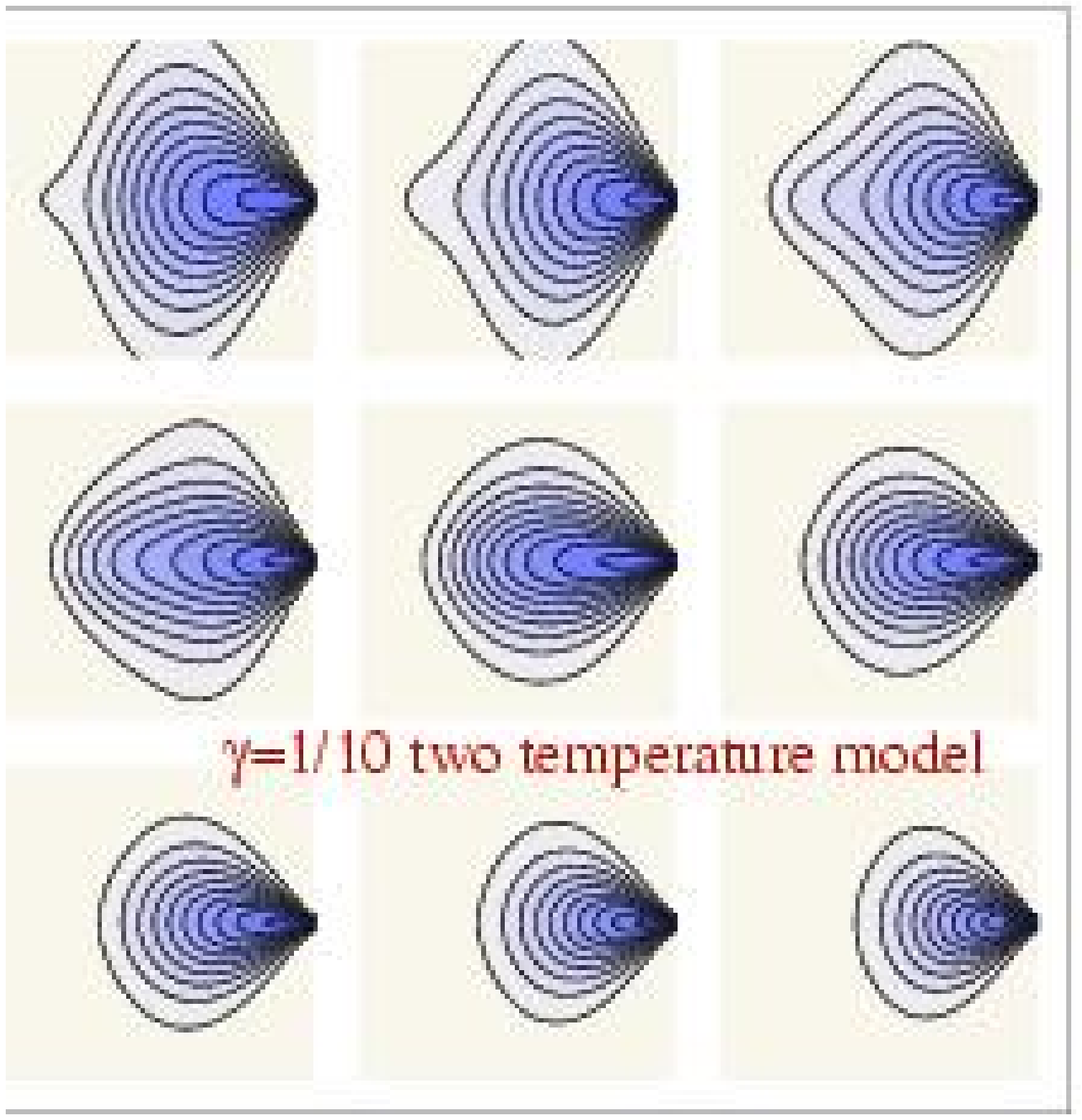}}}
\put(4.5,-0){\centerline{\psfig{width=9cm,file=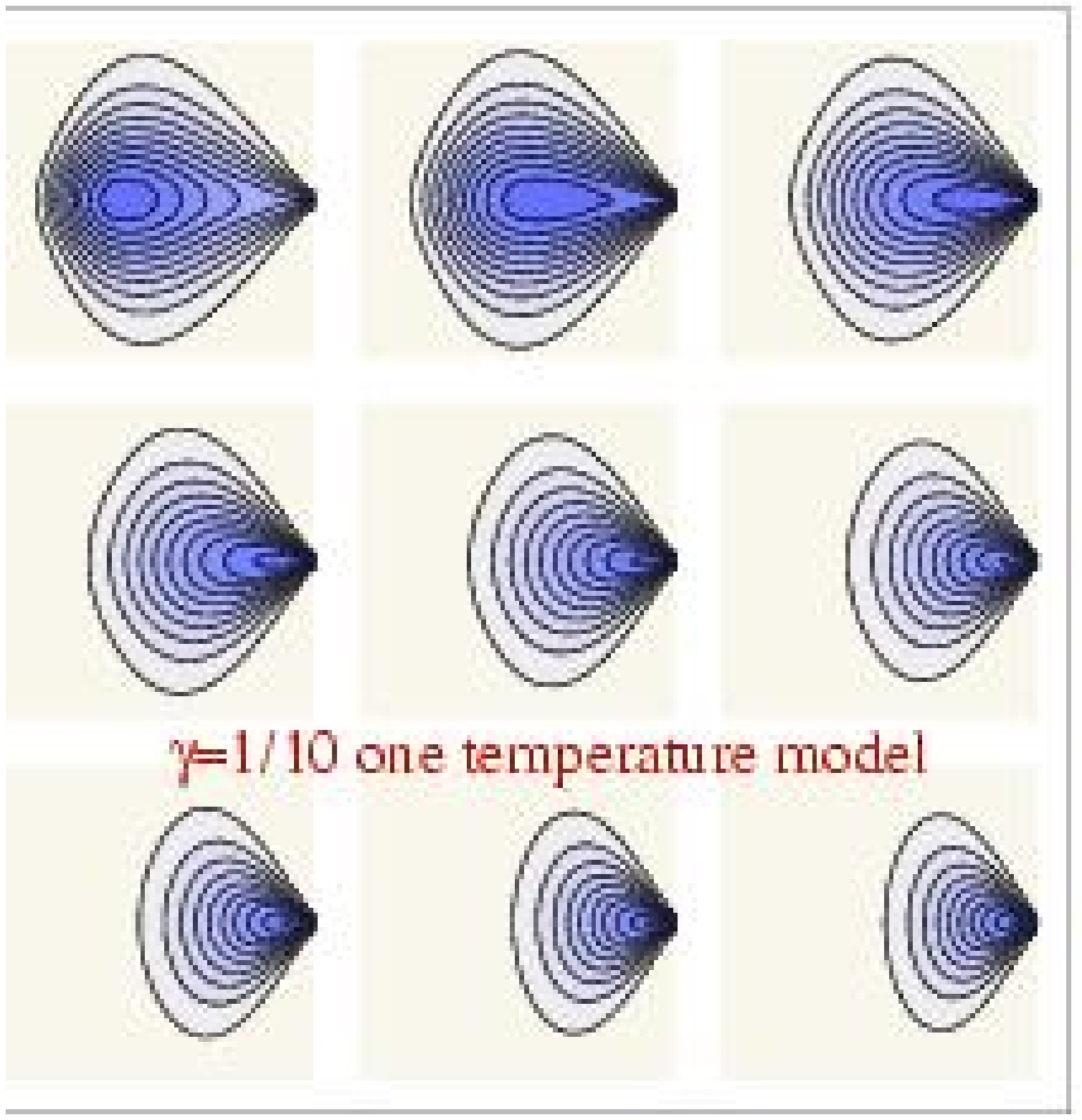}}}
\end{picture}\caption{{\sl In each panel:} Isocontour of $A^\star(b,\mu_b)$
(defined by \Eq{defA}) in the $b,\mu_b  $ plane ($b$ ranging from $-pi/2$ to
$\pi/2$  and $\mu_b$  from $-1$  to  $1$): sections  of increasing  apparent
magnitude  (from  left  to  right   and  top  to  bottom)  {\sl  Top  Left:}
two-temperature models  ($\log\beta =-2$ and  $\log\beta =2$, $\log(L_0)=0$)
{\sl Top Right:} same as top left, but for a unique temperature ($\beta =1$)
model.  The  observed star counts enable  us to distinguish  between the one
and two-temperature  models especially at  the faint end (top  left section)
for significantly non zero $\gamma  \, (=1)$.  {\sl Bottom Left:} Shows that
even the faint end (top left section) of the observed star counts are barely
distinguishable  from the  two-temperature model  ({\sl Bottom  Right:}) for
small $\gamma\, (=1/10)$ .  This demonstrates graphically the requirement to
access  the break  radius, $\ell_0  \propto 1/\gamma$,  of the  potential to
derive statistical distances to the stars.
\label{f:MC}}
\end{figure*}

There is a natural scale $\ell_{0}=(\alpha/\gamma)^{1/(\nu-2)}$ given by the
break in the potential which provides us with a means to lift the degeneracy
between  faint close  stars moving  slowly  and bright  stars moving  faster
farther out.  This  scale reflects the fact that  statistically the dynamics
(i.e. the velocities) gives us a precise indication of distances in units of
$\ell_{0}$.  We can therefore reassign {\it a posteriori} distances to stars
in  the  statistical sense  and  deconvolve  the  colour magnitude  diagram.
\Fig{MC} graphically demonstrates the requirement to access the break radius
of the potential in order to  derive statistical distances to the stars.  It
shows sections of increasing apparent magnitude in the $b,\mu_b$ plane for a
two-temperature model  and for a  one-temperature model (corresponding  to a
unique  absolute  luminosity).   The  observed  star  counts  enable  us  to
distinguish  between the one  and two-temperature  models especially  at the
faint end for significantly non-zero $\gamma$.

Turning back  to \Eq{N22}, it  remains that for  more general $\chi$  it can
still  be inverted  in the  least squares  sense, but  this involves  a less
symmetric  kernel, $K_{1}(x,y  |  u,\beta)$, whose  functional form  depends
explicitly on $\chi$:
\[
K_{1}(x,y | u,\beta)=  \sqrt{\frac{\beta}{2\pi}}\exp \left[ {-\beta u^2
x  -\beta \chi( u  y)} \right]u^3 \, .
\]
The  inversion procedure  which will  be described  in section~\ref{s:simul}
still applies to such kernels.
%
%

\subsection{A Schwarzschild   Model: accounting for the local velocity
ellipsoid anisotropy} \label{s:sch}

Let us now move to more realistic models with a fully triaxial Schwarzschild
Ellipsoid.  Its distribution  function is  given in  terms of  the kinematic
inverse dispersions $\beta_{R},\beta_{\phi}$ and $\beta_{z}$ by:
\begin{equation}
    f_\beta(\M{r},\M{u})=  \sqrt{\frac{\beta_{R}\beta_{z} 
    \beta_{\phi}}{8\pi^{3}}} \exp \left(  {-(\beta_{z}
E_z +\beta_{R}
E_R +\beta_{\phi}
E_\phi  ) } \right) \, ,\EQN{Schwarzschild}
\end{equation}
where 
\begin{equation}
  E_z  =  \frac{1}{2} v_{z}^{2} +\psi_{z}(z)  \,  , \quad  E_R = \frac{1}{2}
v_{R}^{2} \, \quad {\rm and   }\quad  E_\phi = \frac{1}{2}   \left(v_{\phi}-
\bar{v}_{\phi}\right)^{2} \, .  \EQN{energy}
\end{equation}
Here  $\bar{v}_{\phi}$ measures  the mean azimuthal   velocity  in the local
neighbourhood    (which    is  assumed     not to    depend    on  $\beta$),
$\M{V}=(v_{R},v_{\phi},v_{z})$ are  respectively the  radial, azimuthal  and
vertical velocities of a given star  measured in a direct cylindrical system
of  coordinates   centred  at  the Galactic   centre.   
%
%
These velocities are given  as a function  of the velocities measured in the
frame of the sun by
\begin{eqnarray}   
    {v_{\Phi }} &=&  \frac{1}{{R}}{{ \left(  r_\odot\sin  (b )\sin  (\ell  )
      {{{u}}_{b  }}  -r_\odot\cos (b )\sin  (\ell  ) {{{u}}_r} - r_\odot\cos
      (\ell ) {{{u}}_{\ell   }} + r\cos  (b )  \left( {{{u}}_{\ell  }} -\sin
      (\ell  ) {{{u}}_{\odot }} \right) +  \left( r_\odot +  r\cos (b ) \cos
      (\ell ) \right)  {v_{\odot }}  \right) }} \,, \EQN{vphi}  \\
      {v_{R  }} &=&
      \frac{1}{R}\left\{  { \left( r\,\cos  (b   ) - r_\odot\,\cos (\ell   )
      \right) \,   \sin (b )\,  {{{u}}_{b   }}  - r_\odot\,\sin   (\ell  )\,
      {{{u}}_{\ell }} -\,\cos (b )\, \left( {r}\,\cos (b ) - {r_\odot}\,\cos
      (\ell ) \right)  \, {{{u}}_r} + }\right.  \nonumber \\ &&
      \quad \quad
      \left.  {  r_\odot\,{u_{\odot }}   -  r\,\cos  (b   ) \cos( \ell   )\,
      {{{u}}_{\odot  }} +  r\,\cos (b )\,\sin  (\ell  )\,  {{{v}}_{\odot }}}
      \right\} 
      \,,\\ v_{z} &=& \sin (b )\,{{{u}}_r} + \cos  (b )\, {{{u}}_{b }}
      + {w_{\odot }} \, , \EQN{vz}
\end{eqnarray}
where 
\begin{equation}
R = \sqrt{r_\odot^2 - 2r_\odot r\,\cos (b )\,\cos (\ell )
+ r^2\cos   (b )^2}
   \quad {\rm and} \quad  z =  r\sin (b ) \,. \EQN{defR}
\end{equation}
$R$  measures  the projected  distance  (in  the  meridional plane)  to  the
Galactic  centre, while $z$  is the  height of  the star.   Here $u_{\odot},
v_{\odot},w_{\odot}$ and $r_{\odot}$ are  respectively the components of the
Sun's velocity and its distance to the Galactic centre.
The   argument of  the exponential    in  \Eq{Schwarzschild} is a  quadratic
function in $u_{r}$ via \Eqs{vphi}{vz}, so the integration over that unknown
velocity component is straightforward.

In short, we show in  Appendix~\ref{s:slices} that \Eq{N0} has solutions for
families  of distribution  obeying \Ep{Schwarzschild}.   Those  solution are
unique and can  be made explicit for a number of  particular cases which are
discussed there.   They are shown to  be formally equivalent  to those found
for \Eq{ansatzPP}.  For instance, at large distance from the Galactic centre
($r_\odot \!\rightarrow  \!  \infty$) \Eq{N1} along the  plane $\mu_b=0$ can
be re-casted into
\begin{equation}
L^2 \cos(b) A_{2}[b,\ell,\mu_b =0 ,L]=
 {\int \!\! {\int   \sqrt{\frac{\beta}{2\pi}} {\Phi\left[   u^2,\beta
\right]}}\exp \left( -\beta  u^2 x_{3}     - \beta z_{2} 
  \right)u^3 \d u \d\beta } \,,
\quad {\rm with}\quad
x_{3} = 
  \alpha \frac{\sin^2 (b)}{L}  \,, \EQN{N25mt}
\end{equation}
and
\begin{equation}
z_{2} = {{ ( w_\odot\cos b-[{v_\odot}-{\bar v}_{\phi}] \sin b \sin \ell)^2}
 \over      { 2\cos^{2}(b)+  2\sin^{2}(b)\left(       \xi_{R}\cos^{2}(\ell)+
 \xi_{\phi}\sin^{2}(\ell)\right)}}    \,, \quad  \xi_{R}  =    \frac{\beta_{z}}{\beta_{R}}\,  ,   \quad   \xi_{\phi} =
\frac{\beta_{z}}{\beta_{\phi}}\,,  \EQN{defz2}
\end{equation}
which  is of  the  form described  in  section~\ref{s:unique} with  $\nu=0$,
$x_{3}$ replacing $x$ and $z_{2}$ replacing $y$.  We the exeption of the
special 
cases also described in Appendix~\ref{s:slices}, the
solution  can  be  found  via  $\chi^{2}$ minimisation  as  shown  below  in
section~\ref{s:simul}.

%

\subsection{ Epicyclic Model : Accounting for density gradients}
\label{s:epi}

The  above models  do not  account for  any density  or  velocity dispersion
gradients,  which is  a  serious practical  shortcoming.   Let us  therefore
construct an epicyclic model for which the radial variation of the potential
and the kinematic properties of the Galaxy are accounted for.

A distribution function solution of Boltzmann equation with two integrals of
motion (energy and angular momentum) can be writen according to Shu (1969) as
\begin{equation}
    f_{\beta}(\M{r},\M{u})=\Theta(H)\, \frac{\Omega \, \beta^{3/2} \, \rho_D}{\sqrt{2}\,
    \pi^{\frac{3}{2}}\,\kappa \, \sigma^2_R \sigma_z} \exp \left( 
    -\beta \frac{  E_R-E_c} {\sigma^2_R}
    -\beta \frac{E_z}{\sigma^2_z}\right) \, ,
\EQN{deff}
\end{equation}
where $\Theta$ is the Heaviside function while
\begin{equation}
	\Omega = \frac{\kappa }{{\sqrt{2\ \alpha +2}}} \,,\quad
	{{\rho }_D}=  \rho_{\odot } \,\exp \left(\frac{R_{\odot}-R_c}{R_{\rho
	}}     \right) \,, \quad {\rm with} \quad
	R_c= {H^{\frac{1}{\alpha +1}}}\ R_{\odot }^{\frac{\alpha }{\alpha +1}}
        \ V_{\odot }^{-\frac{1}{\alpha +1}} \,,\quad
\end{equation}
$\alpha$  being  the slope  of  the  rotation  curve, $\Omega$  the  angular
velocity, $\kappa$ the epicyclic  frequency, $\rho_D$ the density, $R_c$ the
radius  of the  circular orbit  of  angular momentum  $H$, $\sigma^2_R$  and
$\sigma^2_z$ the square  of the radial and vertical  velocity dispersion and
$\beta$ the kinematic index
\begin{equation}
\sigma^2_R= \sigma_{R_\odot }^{2} \,\exp \left(\frac{2 R_{\odot}-2R_c}
{R_{\sigma_R}}\right) \,,\quad
	\sigma^2_z=  \sigma_{z_\odot }^{2} \,\exp \left(\frac{2 R_{\odot}-2R_c}{   R_{\sigma_z}	}\right) \,,\quad
	E_c= \frac{ \alpha+1}{2 \alpha}  \, H^{\frac{2 \alpha }{\alpha +1}} R_\odot^{-\frac{2 \alpha }{\alpha +1}} V_\odot^{\frac{2}{\alpha +1}} \, .
\EQN{defE}
\end{equation}
Here $\rho_{0}, \Omega, \kappa, \sigma_R,\sigma_{z}$ and $E_{c}$ are
known functions of momentum $H$ given by
\begin{equation}
 H =  r_\odot\cos (b )\sin (\ell ) {{{u}}_r} - r_\odot\sin (b )\sin (\ell )
   {{{u}}_{b }} - \left( r\cos  (b ) - r_\odot\cos  (\ell ) \right) {u_{\ell
   }}  + r    \cos(b)  \cos(\ell)  u_\odot +  \left(r_\odot    - r   \cos(b)
   \cos(\ell)\right) v_\odot \,.\EQN{defL}
\end{equation}  
In the case of a separable potential given by
\begin{equation}
\psi(R,z)= \psi_R(R)+ \psi_z(z)
	 \,,\quad {\rm where} \quad
\psi_R(R)= \frac{R^{2 \alpha } V_{\odot }^{2}
	R_{\odot }^{-2 \alpha }}{2 \alpha } \,,\quad
\psi_z(z)= \frac{1}{2\pi G}\left(\Sigma_0 \left[\sqrt{z^2+D^2}-D\right]+ \rho_{\rm eff} z^2\right) \,,\EQN{defpsi}
\end{equation}
where  G is  the universal  gravity constant,  while $\Sigma_0$,  $\rho_{\rm
eff}$ and $D$ are constants, the energies $E_{z}$ and $E_{R}$ obey
\begin{eqnarray}
  {E_{z }} &=& \frac{\left( \sin (b )\,{{{u}}_r}  + \cos (b )\, {{{u}}_{b }}
   +  {w_{\odot   }}   \right)^2}{2}+\psi_z(z) \,,  \\    {E_R}  &=&
   \frac{{\left( \cos  (b )\, {{{u}}_r} - \sin  (b )\,  {{{u}}_{b }} \right)
   }^2 +  {{u_{\ell }}}^2  +   {{u_{\odot }}}^2  + {{v_{\odot   }}}^2   }{2}
   +\psi_R(R) \nonumber - \sin (\ell  )\, \left( {{{u}}_{\ell }}\, {u_{\odot
   }} -\left(  \cos (b  )\, {{{u}}_r} -  \sin  (b  )\, {{{u}}_{b }}  \right)
   \,{{{v}}_{\odot }} \right)  \nonumber \\&+&  \cos (\ell )\,\left(  \left(
   \cos (b )\, {{{u}}_r} - \sin (b )\,  {{{u}}_{b }} \right) \,{{{u}}_{\odot
   }} + {{{u}}_{\ell }}\, {v_{\odot }} \right) \,,
\end{eqnarray}
while $R$ and $z$ are given by  
\begin{equation}
R = \sqrt{r_\odot^2 - 2r_\odot r\,\cos (b )\,\cos (\ell )
+ r^2\cos   (b )^2}
   \quad {\rm and} \quad  z =  r\sin (b ) \,. \EQN{defR}
\end{equation}

Note that the integration over $ u_{r}$ in \Eq{N0} must now be carried
numerically since $\rho_{0}, \Omega, \kappa, \sigma_\parallel,\sigma_{z}$ are
all functions of $u_{r}$ via \Eq{defL}.

\begin{figure*}\unitlength=1cm
\begin{picture}(14,8)
\put(-7,0.5){\centerline{\psfig{width=7cm,file=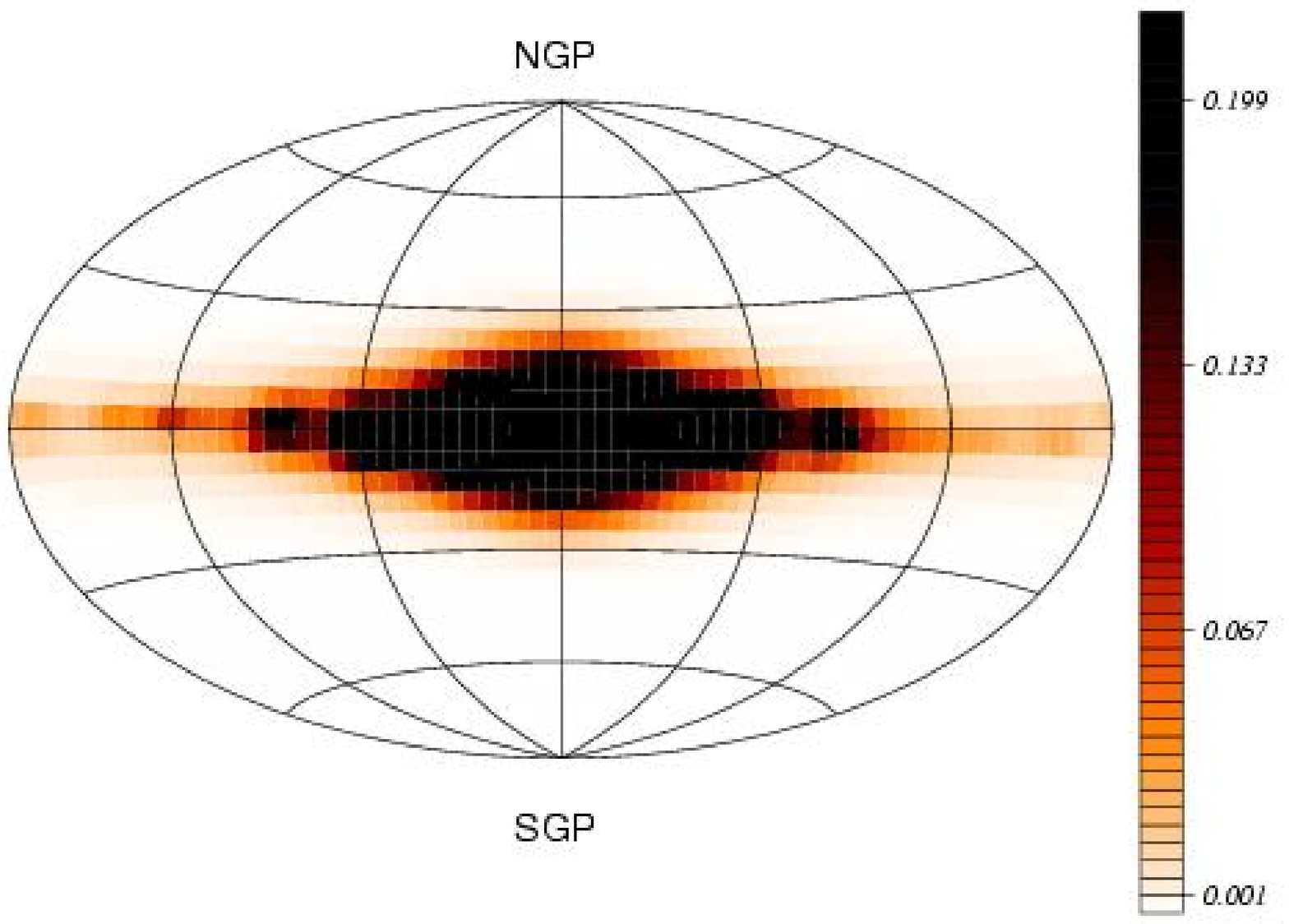}}}
\put(2.5,0.5){\centerline{\psfig{width=7cm,file=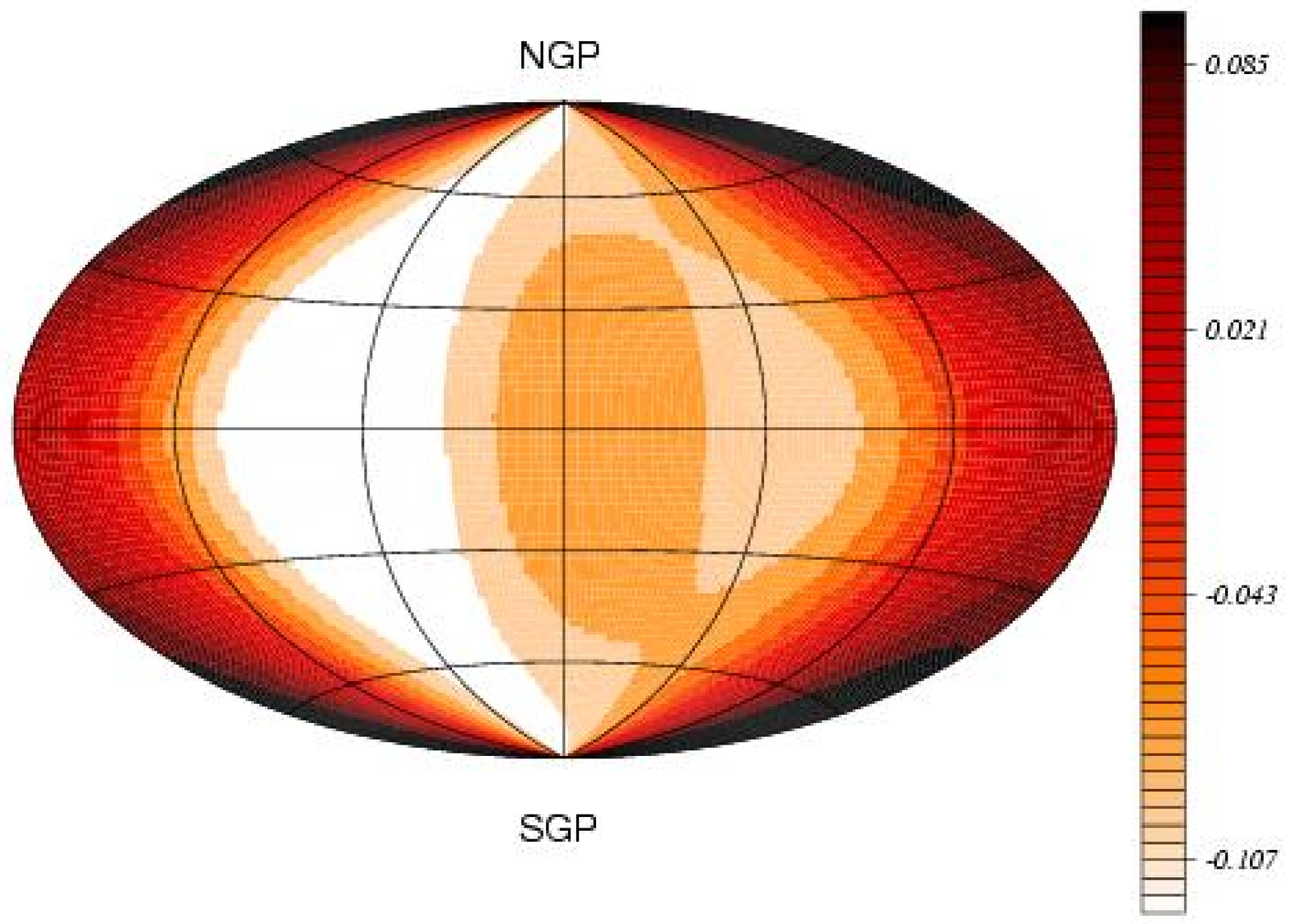}}}
\end{picture}\caption{ {\sl Left:} Aitof projection of the normalised density
distribution for the  epicyclic Shu model. {\sl Right:}  Distribution of the
maximum of  the proper motion along  the galactic longitude  (in ''/yr). The
Galactic center is  in the center of the plot,  longitude is increasing from
the center to  the left. The asymmetry along  the Galactic longitude derives
from the peculiar motion of the sun.
\label{f:shu_distri}}
\end{figure*}

This model, based on the epicyclic theory, accounts for density and velocity
dispersion gradients and is  therefore more realistic than the Schwarzschild
ellipsoid   model  presented   in   Subsection~\ref{s:sch}.   {The   density
distribution together  with the  distribution of the  maximum of  the proper
motion along  the $\ell$ coordinate  are presented figure~\ref{f:shu_distri}
projected onto  the sphere.  The asymmetry along  the Galactic  longitude is
produced by the solar motion.}

\section{Simulations}
\label{s:simul}

\subsection{Method } 
\label{s:Method}

We have chosen to implement a non parametric inversion technique to invert
\Eq{N0}  or  \Ep{N22}.
The  non-parametric inversion  problem is  concerned with  finding  the best
solution  to \Eq{N0}  or  \Ep{N22} for  the  underlying luminosity  function
indexed by  kinematic temperature when only discrete  and noisy measurements
of $A(b,\mu_b,L)$ are available  (e.g.  Dejonghe 1993, Merritt, 1996, Pichon
\& Thi\'ebaut 1998, Lucy 1994, Fadda 1998 and references therein) and most importantly when we have
little prejudice  regarding what  the underlying luminosity  function should
be.  In short, the non-parametric  inversion corresponds to model fitting in
a regime where  we do not want to impose (say  via stellar evolution tracks)
what the  appropriate parametrization of the  model is.  It  aims at finding
the  best compromise between  noise and  bias; in  effect it  correlates the
parameters  so as  to provide  the smoothest  solution amongst  all possible
solutions compatible with a given likelyhood.

An   optimal  approach  should  involve     a maximum likelihood    solution
parametrised  in  terms of the  underlying  6-dimensional distribution.  In
practice such an  approach turns out  to be vastly too  costly for data sets
involving   $10^{6}$  measurements. Binning  is    therefore applied to  our
ensemble of $(\ell,b,\mu_\ell,\mu_b,L$,B-V) measurements.

\subsubsection{ Non-parametric inversion} 
\label{s:NP}

%
The non parametric  solutions of \Eq{N22} and \Ep{N4}  are then described by
their projection onto a complete basis of p $\times$ p functions
 $$\{e_k(\zeta)   e_l(\beta)\}_{  k=1,\ldots,p   \  l=1,\ldots,p}$$   of  finite
(asymptotically  zero) support, which  could be  cubic B-splines  (i.e.  the
unique  $C^2$ function  which  is defined  to  be a  cubic  over 4  adjacent
intervals and  zero outside, with the  extra property that  it integrates to
unity over that interval) or Gaussians:
\begin{equation}
\Phi(\zeta,\beta)  =  \sum_{k=1}^{p} \sum_{l=1}^{p}   \Phi_{kl}  \,  \, e_{k}(\zeta)
e_{l}(\beta) , \EQN{decomp}
\end{equation}
The parameters  to  fit  are  the  weights $\Phi_{kl}$.
Calling  $\M{x}=\{ \Phi_{kl}\}_{k=1,..p,l=1,..p}$
(the   parameters)  and 
$\tilde{\M{y}}=\{ L^2 \cos(b) A(x_{i},y_{j})\}_{i=1,..n,j=1,..n}$ 
(the n $\times$ n  measurements with $L^2 \cos(\beta)$ a function of $x_i,y_i$
via \Eq{defx}),   \Eq{N22} 
then becomes formally
\begin{equation}
\tilde{\M{y}}= \M{a} \cdot \M{x} \, , \EQN{yax}
\end{equation}
where $\M{a}$ is a $(n,n)\times (p,p)$ matrix with entries given by
\begin{equation}
 a_{i,j,k,l}=  \left\{  {\int \!\!  {\int \sqrt{\frac{\beta}{2\pi}} {e_{k}(\zeta^{2}) e_{l}(\beta) }}\exp
\left[  {-\beta \zeta^2 x_{i} -\beta  \chi( \zeta y_{j})} \right]\zeta^3  \d \zeta \d\beta }
\right\}_{i,j,k,l}\, .  \EQN{eqnNP}
\end{equation}

For the epicyclic model the measurements are 
$\tilde{\M{y}}=\{A_{ijklm}=
A(\ell_{i},b_{j},\mu_{\ell_k}\mu_{b_l},L_{m})\}_{i=1,..n_{1},
j=1,..n_{2},,k=1,..n_{3},,l=1,..n_{4},m=1,..n_{5}}$   
and  $\M{a}$  is  a  $(n_{1},n_{2},n_{3},n_{4},n_{5})\times (p_{1},p_{2})$
matrix with entries given by
\begin{equation}
 a_{i,j,k,l,m,q,s}= \left\{ {\int  \!\!  \int  \!\! {\int {e_{q}(L_m  r^{2})
 e_{s}(\beta)  }} f_{\beta}(\ell_{i},b_{j},\mu_{\ell_k}\mu_{b_l},r,u_r)  r^4
 \d r \, \d u_{r} \, \d\beta } \right\}_{i,j,k,l,m,q,s}\, , \EQN{eqnNP}
\end{equation}
with $f_{\beta}$ given by \Eq{deff}.

Assuming  that we  have access  to  discrete  measurements of $A_{ij}$  ( or
    $A_{ijklm}$ via
binning  as discussed   above),  and  that  the noise  in   $A$ can  be
considered to be Normal, we can estimate the error between the measured star
counts and the non-parametric  model by
\begin{equation}
\R{L}_\R{}(\M{x}) \equiv \chi^2(\M{x}) = 
		\T{({\tilde{\M{y}}} - \M{a}\mdot \M{x})} \mdot \M{W}
		\mdot ({\tilde{\M{y}}} - \M{a} \mdot \M{x}) \,, \EQN{Lquad}
\end{equation}  where 
the weight  matrix $\M{W}$ is  the inverse of  the covariance matrix  of the
data (which is diagonal for  uncorrelated noise with diagonal elements equal
to one over the data variance).

The  decomposition in  \Eq{decomp} typically  involves many  more parameters
than  constraints,  such that  each  parameter  controls  the shape  of  the
function  only   locally.   The  inversion  problem   corresponding  to  the
minimization  of \Eq{Lquad} is  known to  be ill-conditioned:  Poisson noise
induced by the very finite sample of stars may produce drastically different
solutions  since these  solutions  are  dominated by  artefacts  due to  the
amplification of noise.  Some trade-off must therefore be found between the
level of  smoothness imposed  on the  solution in order  to deal  with these
artefacts on the one hand, and the level of fluctuations consistent with the
amount of  information in the  data set on  the other hand.  Finding  such a
balance is  called the  ``regularization'' of the  inversion problem  and in
effect implies that between  2 solutions yielding equivalent likelihood, the
smoothest  is  chosen.  In  short,  the solution  of  \Eq{yax}  is found  by
minimizing the quantity
\[Q(\M{x})=L(\M{x})+\lambda\,R(\M{x})\] where $L(\M{x})$ and $R(\M{x})$ are,
respectively, the  likelihood and  regularization terms given  by \Eq{Lquad}
and
\begin{equation}
 	\R{R}_\R{}(\M{x}) = \T{\M{x}} \mdot \M{K} \mdot \M{x}\,, \EQN{Pquad}
\end{equation} where $\M{K}$ is a positive definite matrix, which is chosen 
so that R in
\Eq{Pquad} should be non-zero when $\M{x}$ is strongly varying as a function
of its indices.  In practice, we use here
\[
\M{K}=
 \M{K}_{3} \otimes \M{I} +  \M{I} \otimes \M{K}_{3} +
 2 \M{K}_{2} \otimes \M{K}_{2}\,,
\]
where $ \otimes  $ stands for  the outer product, $  \M{I} $ is the identity
and      $\M{K_{2}}  = \T{\M{D_{2}}}\cdot   \M{D_{2}}$     \,, $\M{K_{3}}  =
\T{\M{D_{3}}}\cdot \M{D_{3}}$. Here  $\M{D}_{2}$ and  $\M{D}_{3}$ are finite
difference second order operators (of dimension $(p-2) \times p$ 
and $(p-3) \times p$
respectively) defined by
\begin{equation}
    	\M{D}_{2}=
{\rm Diag}_{2}[-1,2,-1] \equiv  \left[\begin{array}{cccccc}
	-1&2&-1&0&0& \ldots \\
	0&-1&2&-1&0& \ldots \\
	0&0&-1&2&-1 & \ldots\\
	0&0&0&-1&2 & \ldots\\
	\ldots&\ldots&\ldots&\ldots&\ldots &\ldots 
\end{array}\right]\ ,
\,\,
	\M{D}_{3}=
{\rm Diag}_{3}[1,-3,3,-1] \equiv  \left[\begin{array}{cccccc}
	1&-3&3&-1&0& \ldots \\
	0&1&-3&3&-1& \ldots \\
	0&0&1&-3&3 & \ldots\\
	0&0&0&1&-3 & \ldots\\
	\ldots&\ldots&\ldots&\ldots&\ldots &\ldots 
\end{array}\right]  
 \, ,  
\end{equation}

This choice corresponds a quadratic operator whose kernel include planes and
paraboloids.   The operator  $\M{K}$ is  typically non  zero  (and therefore
penalizes the minimization of $Q(\M{x})$) for unsmooth solutions (i.e. those
leading to strong variations in the coefficients $\Phi_{kl}$).

The  Lagrange  multiplier  $\lambda>0$  allows  us  to  tune  the  level  of
regularization.  The introduction  of the  Lagrange multiplier  $\lambda$ is
formally justified by the fact  that we want to minimize $Q(\M{x})$, subject
to   the    constraint   that   $L(\M{x})$   should   be    in   the   range
$N_\R{data}\pm\sqrt{2\,N_\R{data}}$). In practice, the minimum of
\begin{equation}
	Q_\R{}(\M{x}) =
	\T{({\tilde{\M{y}}} - \M{a}\mdot \M{x})} \mdot \M{W}
	\mdot ({\tilde{\M{y}}} - \M{a} \mdot \M{x})
	+ \lambda \, \T{\M{x}} \mdot \M{K} \mdot \M{x} \, 
	\label{e:Q-quad}
\end{equation}

\noindent   is:

\begin{equation}
	\M{x}_\R{} = (\T{\M{a}} \mdot \M{W} \mdot \M{a} + \lambda \, \M{K})^{-1}
	\mdot \T{\M{a}} \mdot \M{W} \mdot \tilde{\M{y}} \,.
\label{e:Q-quad-solution} 
\end{equation}

The last remaining issue involves  setting the level of regularization.  The
so-called  cross-validation  method  (Wahba   1990)  adjusts  the  value  of
$\lambda$ so  as to minimize residuals  between the data  and the prediction
derived from the data.  Let us define

\begin{equation}
	\tilde{\M{a}}(\lambda) =\M{a} \mdot (\T{\M{a}} \mdot \M{W} \mdot \M{a} +
	\lambda\,\M{K})^{-1} \mdot \T{\M{a}} \mdot \M{W} \,.
\label{e:Q-quad-lambda}
\end{equation}

We make use of the value for $\lambda$ given by Generalized Cross Validation
(GCV) (Wahba \& Wendelberger 1979) estimator corresponding to the minimum of

\begin{equation}
	\lambda_0 \equiv GCV(\lambda) = {\rm min}_\lambda\left\{
	\frac{||( \M{1} - \tilde{\M{a}}) \mdot \tilde{\M{y}} ||^2}{
	\left[{\rm trace}( \M{1} - \tilde{\M{a}}) \right]^2} \right\} \,.
\EQN{GCV}
\end{equation}

Note that  the model \Eq{yax} is  linear and so is \Eq{Q-quad-solution} but
this need not be the case when positivity is required. We would then resort
to non linear minimization of \Eq{Q-quad}.

\subsubsection{ Positivity}
\label{s:positivity}

When  dealing with  noisy datasets,  the non-parametric  inversion technique
presented  above (section~\ref{s:NP}) may  produce negative  coefficients in
the  reconstructed luminosity  function.  In  order to  avoid  such effects,
positivity can be imposed  on those coefficients $\Phi_{kl}$ in \Eq{decomp}.
A simple  way to achieve positivity  is to use an  exponential transform and
introduce $\varphi$ so that :
\begin{equation}
 \Phi \equiv \Phi_0 \exp \left( \varphi \right)\, , \EQN{defvarphi}
\end{equation}
where  $\Phi_0$ corresponds  to our  first  guess for  $\Phi$ (here  $\Phi_0
\equiv 10^3$).  A first order Taylor expansion of \Eq{defvarphi}, together 
with \Eq{yax} yields
\begin{equation}
	\tilde{\M{y}}' \equiv \tilde{\M{y}} -\M{a} \mdot \Phi_0 =\M{a} \mdot \Phi_0 
	\mdot \varphi \equiv \M{a} \mdot \M{x}' \, ,
\EQN{positivity}
\end{equation}
which   defines  $\tilde{\M{y}}'$   and  $   \M{x}'  $.   We   first  invert
\Eq{positivity}  for $ \M{x}'  $.  The  algorithm is  then iterative  and we
invert in turn for $ \M{x}_n'$
\[
\tilde{\M{y}'}_n= \M{a} \cdot \M{x}_n' \, , \quad {\rm where}\quad
 \tilde{\M{y}'}_n= \tilde{\M{y}} -\M{a}\mdot \Phi_{n-1}  \quad {\rm and} 
\quad {\M{x}'}_n=   \Phi_{n-1} \mdot \varphi_n \EQN{yax'} 
\]
 the luminosity function is expressed as
\be
 \Phi_n =  \Phi_{n-1} \exp \left( \eta_{\rm conv} \M{x}'_n / 
\Phi_{n-1} \right) \EQN{defetaconv}
 \ee
 in \Eq{positivity} for the iteration number $n$.
{ In  practice convergence is  controlled via a parameter,  $\eta_{\rm conv}
\in [0,1]$, which  fixes the amplitude of the  correction in \Eq{defetaconv}
in order to remain within the r\'egime of the Taylor expansion.  } It should
be   emphasized   that   using   equation  \Eq{positivity}   together   with
\Eq{Q-quad-solution} (replacing $\M{x}$ by  $\M{x'}$) does not lead directly
to  the expected  luminosity function  but to  a correction  that has  to be
applied  to $\Phi_0$.   \\ 

We will  now proceed to invert  \Eq{N2} in two  r\'egimes: the Schwarzschild
model  described by  \Eq{Schwarzschild} and  the epicyclic  model,  given by
\Eq{deff}.   The former  model is  dimensionally less  demanding,  while the
latter  is  more  realistic  since  it accounts  for  density  and  velocity
dipsertion gradients.

\subsection{ Simulated Schwarzschild models} 
\label{s:simsch}

\begin{figure*}\unitlength=1cm
\begin{picture}(14,7)
\put(-5,-0.5){\centerline{\psfig{width=11cm,file=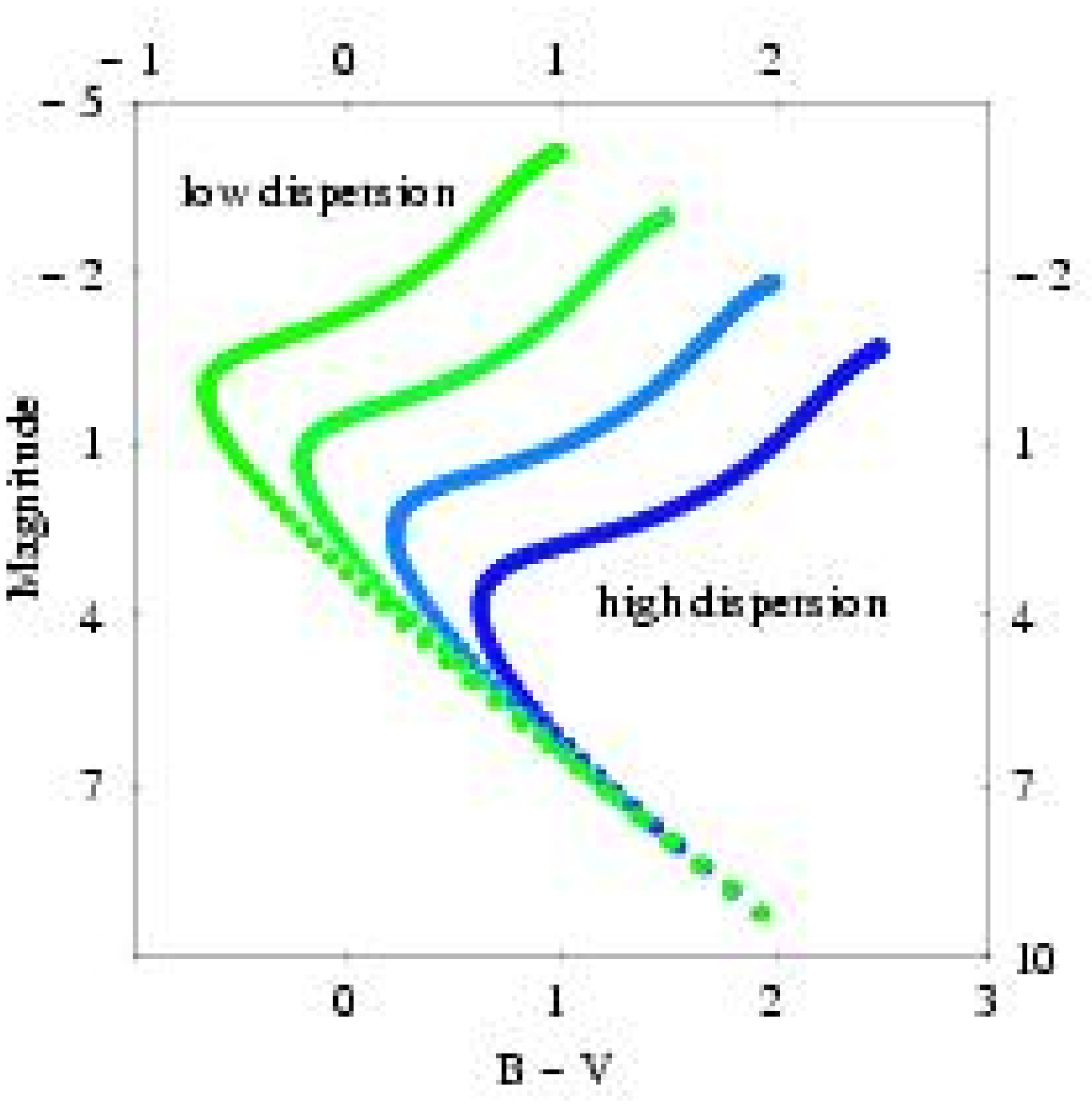}}}
\put(2.5,-0.5){\centerline{\psfig{width=7.5cm,file=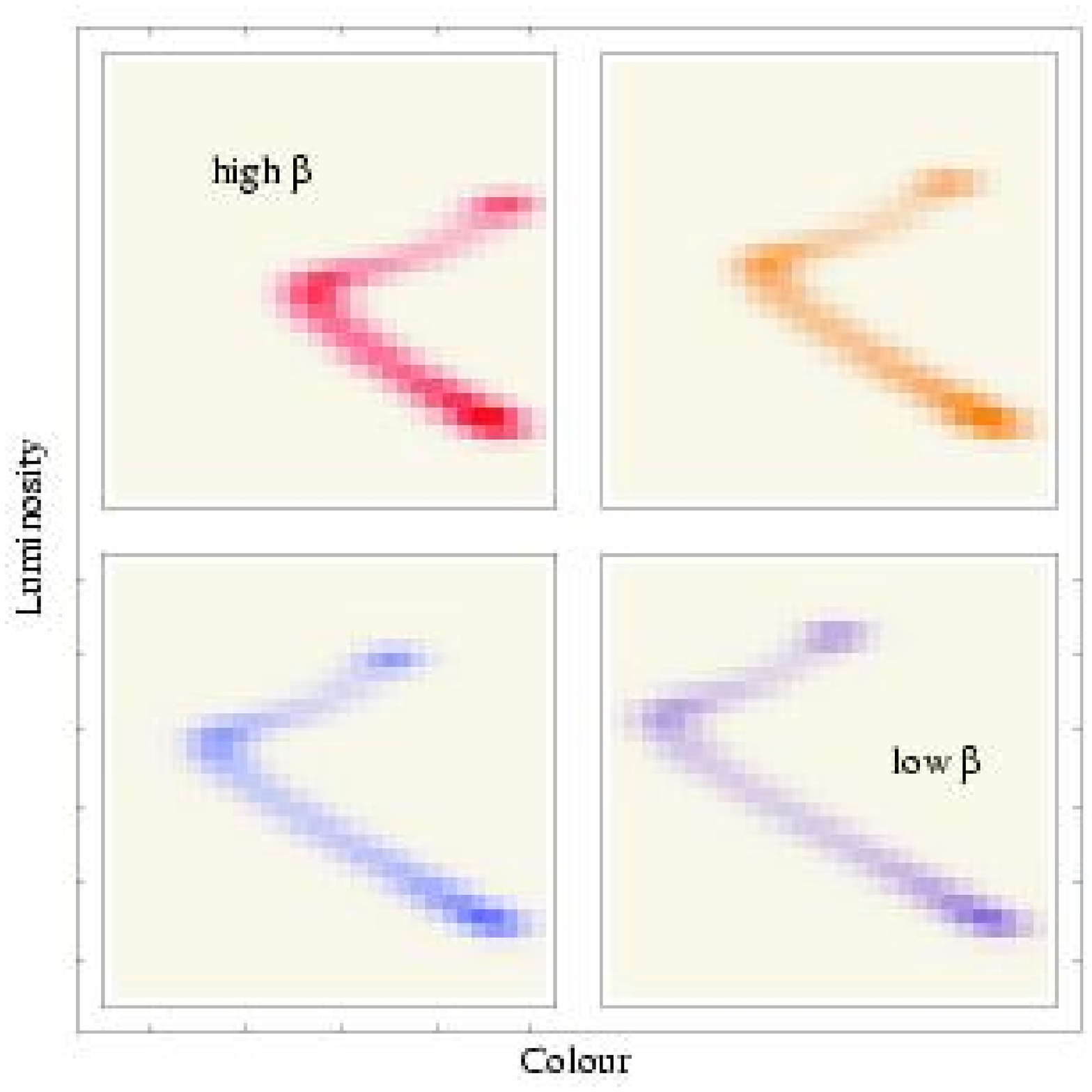}}}
\end{picture}\caption{ {\sl   Left:} Fictitious tracks corresponding to 
increasing    (from left to    right)  kinematic  temperature  {\sl  Right:}
decomposition  of  corresponding colour  magnitude    diagram into its  four
components, weighted  by the IMF  on each track.  The  image in the observed
plane $(\mu_b,b,L$,{\rm B-V}) of these tracks is shown in
\Fig{obs}.
\label{f:track}}
\end{figure*}

\begin{figure*}\unitlength=1cm
\begin{picture}(14,14)
{\centerline{\psfig{width=13cm,file=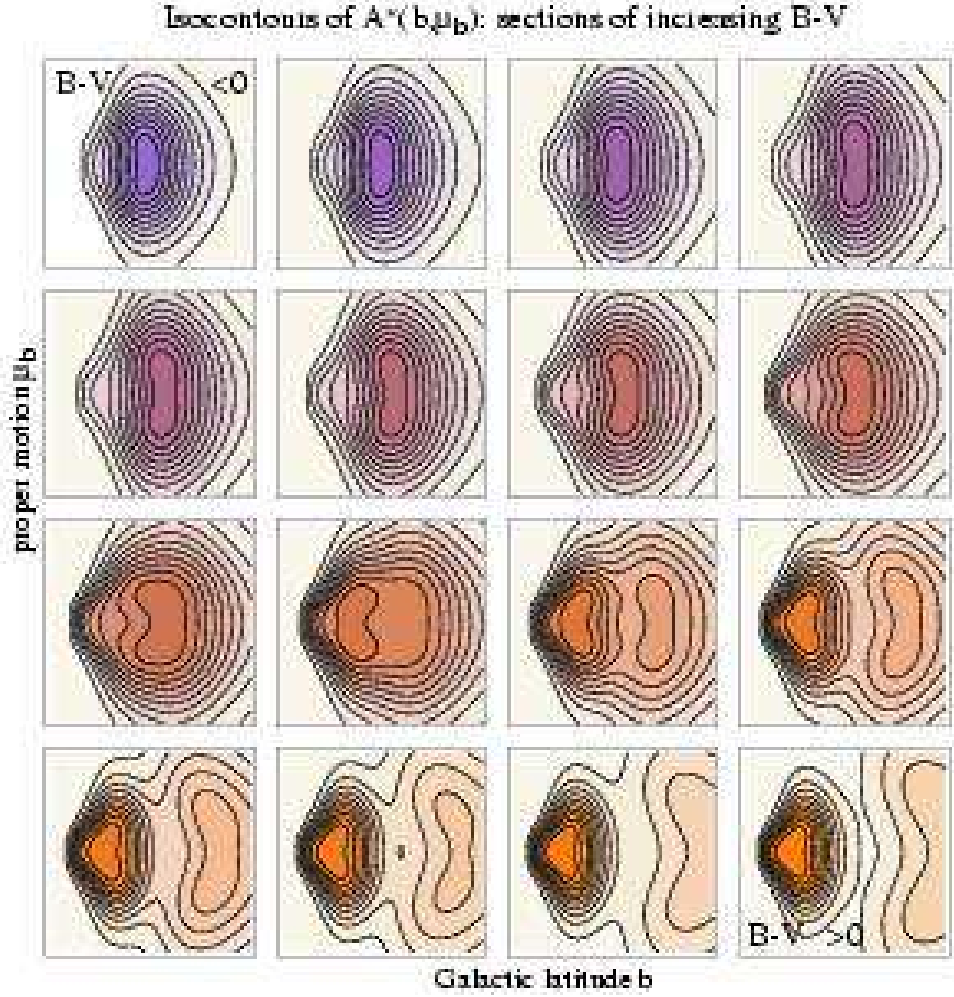}}}
\end{picture}\caption{ $A^\star(b,\mu_{b}$,B-V) in the $b,\mu_b  $  plane  for
increasing B-V (from  left to  right   and top to  bottom)   at a fixed  apparent
magnitude   $L=1/10$  of    the   model   described  in   \Fig{track}.
Interestingly,  the   multiple kinematic   components  of the   redder
sections  display distinct extrema for   opposite values of $\mu_b$ at
fixed Galactic  latitude, $b$, and also as  a function of  $b$ at fixed
proper motions.
\label{f:obs}}
\end{figure*}

We will  first focus on the  inversion of \Eq{N22}, rather  than \Ep{SN1} or
\Ep{N24} (which  were shown  to be equivalent  in the zero  asymmetric drift
approximation) and \Ep{N25mt} (which was also shown to be of the same form).
Special emphasis  is put on  the toy model described  in section~\ref{s:toy}
while carrying the inversion on a superposition of 4 kinematically decoupled
populations  with distinct  main  sequence turn-off  magnitudes.  These  are
illustrated  on   \Fig  {track}  which   displays  the  4   fictious  tracks
corresponding to  increasing kinematic temperature  weighted by some  IMF on
each track.  The image in the observed plane $(\mu_b,b,L$,\rm{B-V}) of these
tracks is shown in \Fig{obs} which shows isocontours of $A^\star$ defined by
(corresponding to \Eq{defA1} with $c=3/4$)

\begin{equation}
A^\star(b,\mu_b,L)     =A(b,\mu_b,L)    \cos(b) \,\left(   {\frac{\gamma   \
{{\sin}^\nu}(b)\     \big[{{\mu_b      }^2}\   {{\sec}^2}(b)+2\   \alpha   \
{{\sin}^2}(b)\big]}{2\ {{L^{\nu/2-5/3}}}}} \right)^{3/4}\EQN{defA} 
\end{equation}
in  the $b,\mu_b $ plane  for increasing B-V at a fixed apparent magnitude
$L=1/10$. The multiple kinematic components  of the redder sections display
distinct extrema for opposite values of $\mu_b$  at fixed Galactic latitude,
$b$, and also as a function of $b$ at fixed  proper motions.  In all figures
$\gamma$ is chosen equal  to $1$ (unless specified  otherwise) and $\nu$  to
$3$. For simplicity, we also numerically approximate $K_0$ in \Eq{defK} by a
Gaussian since the matrix elements in \Eq{eqnNP} are then analytic.

\subsection{ Simulated epicyclic models} 
\label{s:simshu}

In  order to  test the  inversion procedure,  a set  of 4  HR  diagrams with
different  turn-off  luminosity was  constructed  assuming a  mass-luminosity
relation (MLR)  and a Salpeter  initial mass function (IMF).  The luminosity
function of each population scales like
\begin{equation}
\Phi_0  \propto \mdot L^{-\frac{\varsigma}{\tau}} \, ,
\end{equation}
where $\tau$ (the slope  of the MLR on a logarithmic scale)  was set to 3.2,
which is  caracteristic of the main  sequence, and $\varsigma$  to 2.35 (the
IMF slope).  The scaling factor, fixes  the number of stars in the simulated
galaxy.  The tracks associated with those  HR diagrams were then binned on a
$20 \times  20 \times 4$ grid  in the [$L_0$,$B-V$, $beta$]  space; those HR
diagrams   represent   the    absolute   luminosity   function,   $\Phi_{\rm
B-V}(L_0,\beta)$.   The {\it  observed} counts  were then  computed assuming
that  each  track  corresponds to  a  given  kinematic  index and  that  its
distribution can be reproduced by  the epicyclic model of the same kinematic
index {\sl i.e.}:

\begin{equation}
dN(\ell_i,b_j, \mu_{\ell,k}, \mu_{b,l},L_m,B-V) = 
a_{i,j,k,l,m,q,s} \mdot \left(\Phi_{0}(B-V)\right)_{q,s}  \d \mu_{\ell} \d \mu_{b} \d
{\ell} \cos{b} \d b \d L \d (B-V)\, ,
\end{equation}
where $a_{i,j,k,l,m,q,s}$ is given by \Eq{eqnNP}.
Poisson noise was  introduced in corresponding histograms used  as input for
the inversion procedure.  It should  be emphasized that constructing such HR
diagrams does not challenge the  relevance of our physical model, \Eq{deff},
but only our  ability to recover a given luminosity  function.  The model LF
need not be  very realistic at this stage.  The  parameters of the epicyclic
model  given in  Table~\ref{t:shu} were  set so  as to  reproduce  the local
neighbourhood  according to Bienaym\'e  \& S\'echaud  (1997) and  Vergely et
al. (2001).   Figure~\ref{f:model_shu} show the assumed  and reconstruced HR
diagrams for the four populations  in the [$L_0$,$B-V$] plane for this model
while figure~\ref{f:rec_shu} shows the  reconstruction error in \% for those
two figures.\\
\begin{center}
\begin{table}
\begin{tabular}{|l|l|l|}
\hline
Distribution function & Potential & Solar motion \\
\hline
 $R_\rho = 2.5 \,\, {\rm kpc}$ & $D = 240 \,\, {\rm pc}$ & $R_\odot=8.5 \,\,
{\rm kpc}$\\
 $R_\rho = 2.5 \,\, {\rm kpc}$ & $\Sigma_0 = 48 \,\, M_\odot/{\rm pc}^2$ &
$V_{\rm LSR} = 220 \,\, {\rm km/s}$\\
 $R_{\sigma_R} = 10 \,\, {\rm kpc}$ & $\rho_{\rm eff} = 0.0105 \,\,
M_\odot/{\rm pc}^3$ & $U_\odot = 9. \,\, {\rm km/s}$\\
 $R_{\sigma_z} = 5 \,\, {\rm kpc}$ & $\alpha = -0.1$ & $V_\odot = 5.2 \,\,
{\rm km/s}$\\
 $\sigma_{R_\odot} = 48 \,\, {\rm km/s}$ & & $W_\odot = 7. \,\, {\rm
km/s}$\\
 $\sigma_{z_\odot} = 24 \,\, {\rm km/s}$ & &\\
 $\rho_\odot = 0.081 \,\, M_\odot/{\rm pc}^3$ & &\\
\hline
\end{tabular}
\caption{Parameters   used   for    the   epicyclic   model   described   in
Section~\ref{s:epi}}
\label{t:shu}
\end{table}
\end{center}
%

\section{ Results }

\subsection{ the Schwarzschild models}
\label{s:ressch}

\begin{figure*}\unitlength=1cm
\begin{picture}(14,8)
\put(-2,0){\centerline{\psfig{width=15cm,file=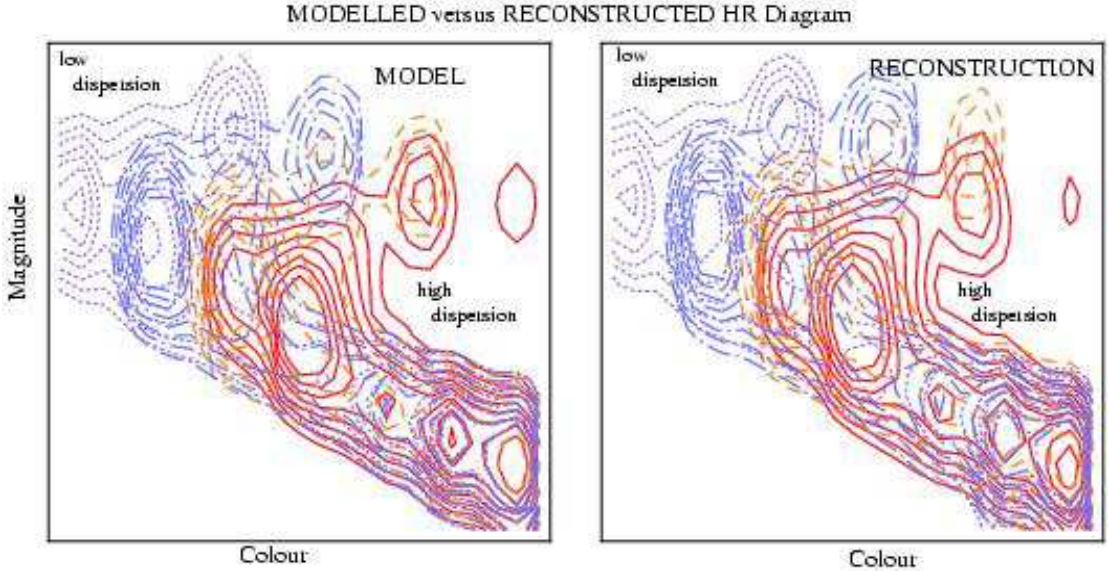}}}
\end{picture}\caption{ {\sl   Left:}  assumed and
{\sl   Right:} reconstructed HR diagram  together  with its decomposition in
kinematic temperature. Note that the wiggly structures are a property of the
model, and are well recovered by  the inversion procedure.  The plain, dash,
dot dash, short dash curves correspond  to the 4 dispersions associated with
the 4 populations   with distinct  main  sequence  turn-off radii shown   in
\Fig{track}.
\label{f:HR}}
\end{figure*}

\begin{figure*}
     \unitlength=1cm
{\centerline{\psfig{width=12cm,file=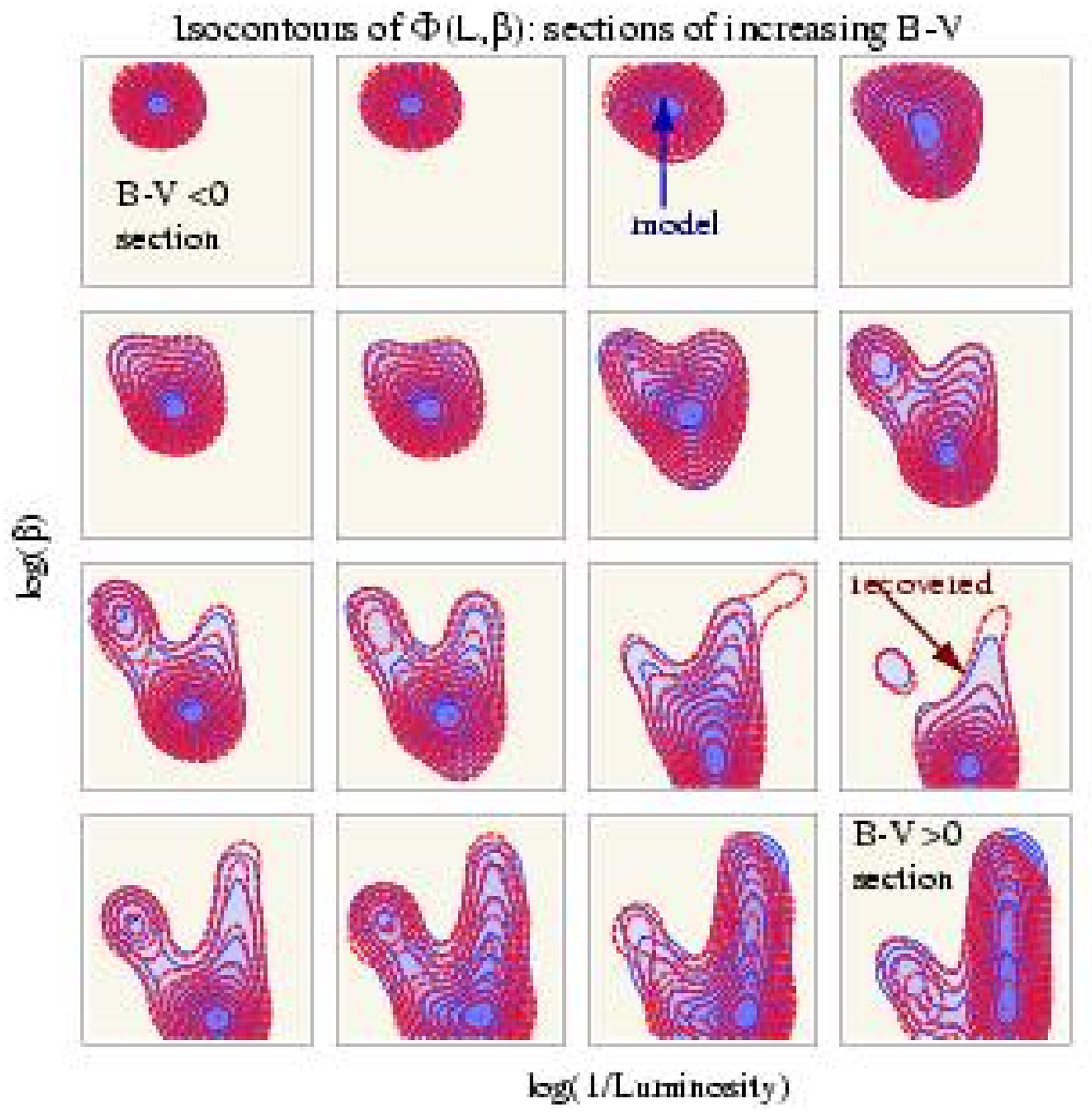}}}
 \caption{ assumed model  (plain line,   filled  contour) and  non  parametric
 deprojection (dashed line)  overlaid on top  of the expected contours in the
 $\left(\log1/L,\log\beta\right)$  plane   for   increasing $B-V$   sections.
 Errors in the deprojection are largest for lower  contours.  Note that these
 sections are orthogonal to those superposed in \Fig{HR}.
 \label{f:HR2}}
 \end{figure*}

The above non-parametric inversion technique was implemented on $19\times 19
\times 19$  data sets (and up to  $41\times 41 \times  41$) corresponding to
measurements in $X,Y, B-V$ (\Eq{defX}).  For  each $B-V$ section, we recover
$19\times 19$ (resp.  $41\times 41$) coefficients $\M{x}_{ij}$ corresponding
to values   of   $U,B$, which  implies that   our  resolution  in  kinematic
dispersion is logarithmic.   \Fig{HR}  shows isocontours of the  assumed and
reconstructed HR  diagram as its decomposition in  kinematic dispersion.  In
this zero noise, no bias r\'egime, the relative discrepancy between the data
and the projection of the model is less than one part in $10^{3}$ while that
between the model and the inversion is lower than $10 \%$ (the corresponding
loss in accuracy  is characteristic of  non-parametric deconvolution).  Note
that the wiggly   structures  are a property    of the model,  and  are well
recovered by  the  inversion    procedure.   \Fig{HR2} shows   the    actual
deprojection overlaid on  top of the expected  contour  of the model in  the
(logarithmic)  $(\beta,L)$ plane  for  increasing    values of $B-V$    (the
projection of the fit in  data space is  not displayed because residuals of
the fit would be  too small to  be seen).   Errors  in the deprojection  are
largest  for lower contours.  Note  that the contours of \Fig{HR} correspond
to sections of  the cube shown in  \Fig{HR2}  which are orthogonal to  those
displayed in \Fig{HR2}.

\subsubsection{ Errors in measurements and finite sample }

The above  results were achieved  assuming infinite numbers   of stars and no
truncation in  apparent magnitude.   The  Poisson  noise induced by the
finite number of stars (for which accurate  photometric and kinematic data is
available), as well as the actual error in those measurements, are likely to
make the inversion of \Eq{N22} troublesome.  

\begin{figure*}\unitlength=1cm
\begin{picture}(14,8)
\put(-5.5,-0.25){\centerline{\psfig{width=10cm,file=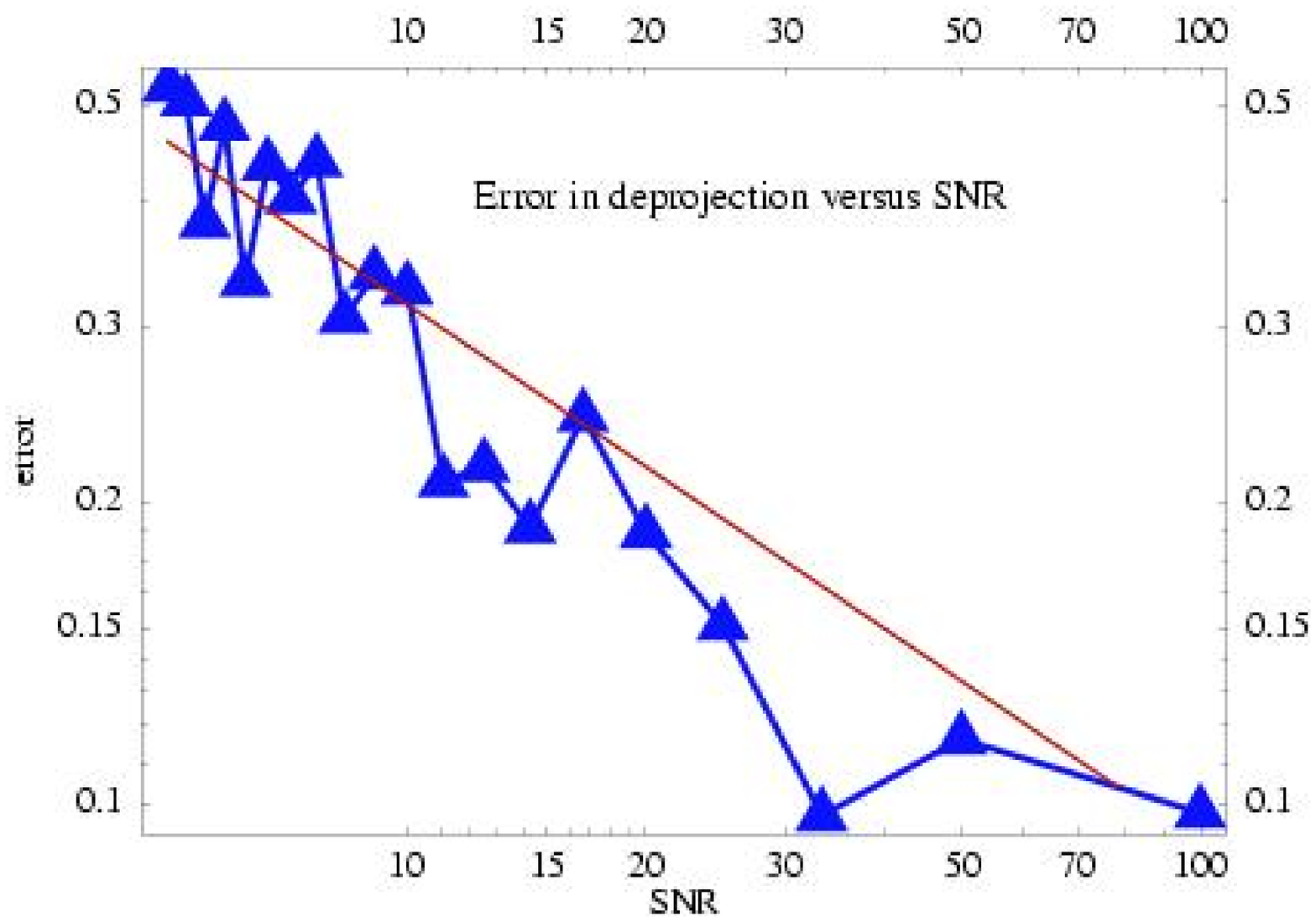}}}
\put(4,-0.){\centerline{\psfig{width=7cm,file=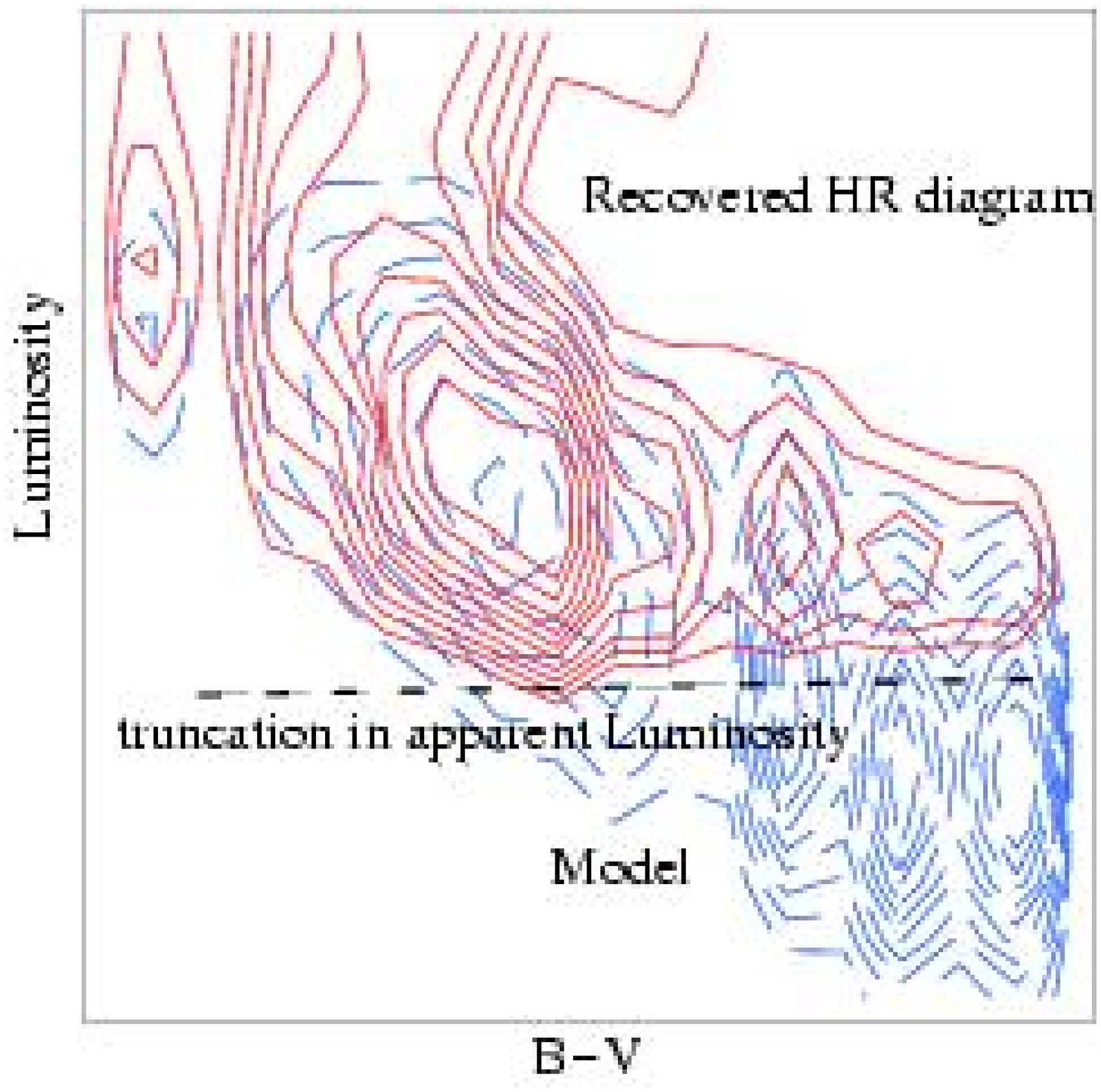}}}
\end{picture}\caption{ {\sl Left Panel:} the mean absolute residual of the
luminosity function,  $\sum_{ij}    |\Phi_{ij}^{\rm   recov}-\Phi_{ij}^{\rm
input}|/\sum_{ij} |\Phi_{ij}^{\rm input}|$ versus the  signal to noise ratio
in logarithmic coordinates. This  graph demonstrates that the non-parametric
inversion sketched in section~\ref{s:NP} is robust  with respect to sampling
or   measurement noise.  {\sl Right  Panel:}    the effect of truncation  in
magnitude  on the main  sequence: plain  line:  recovered HR diagram with  a
truncated data set; dashed  line: recovered  HR diagram without  truncation.
As expected, the truncation in apparent magnitude removes the information at
the bottom of the main sequence.
\label{f:SNR}}
\end{figure*}

\Fig{SNR} shows  how the error  in the recovered  HR diagram decreases  as a
function of  the signal to  noise ratio  in the data  which for the  sake of
simplicity,  was assumed  to be  constant while  the noise  was taken  to be
Gaussian (corresponding to the large number  of stars per bin). Note that in
reality  the  signal  to  noise  ratio will  clearly  be  apparent-magnitude
dependent, and  distance dependent (because of extinction  and proper motion
errors).   \Fig{SNR} also  shows how  the truncation  in  apparent magnitude
induces a truncation in absolute magnitude  (here we truncate in $Y$ since a
truncation  in  $L$ induces  a  truncation  in $Y$  but  none  in $X$  given
\Eq{defX}).
%

\subsection{ the epicyclic models} \label{s:resepi}

{The inversion technique  has been implemented over a $36  \times 9 \times 7
\times 7 \times 10 \times 20 \times  4$ model which correspond to a bin size
projected  onto the sphere  of $10  \times 10$  degrees in  position sampled
linearly, 7  bins in proper  motion ranging from  -0.2 to 0.2 mas/yr  and 20
bins  in apparent and  absolute luminosity  correpsonding to  an integration
over the line of  sigth from 0.1 pc to 4 kpc (those  are also linear bins in
luminosity, which correspond to a  logarithmic binning in radius).  The four
kinematic indexes (ranging  from 0.8 to 120) were set  to reproduce a series
of disks with density scale heights ranging from nearly 200 pc to 1 kpc({\sl
i.e.}   corresponding to  thin and  thick disks).   The mean  SNR  for these
simulation is 2000, ranging from 20  (on the giant branches) to 70000 on the
bottom of the main sequence.\\}

\begin{figure*}\unitlength=1cm
\begin{picture}(14,8)
\put(-5.5,0.5){\centerline{\psfig{width=7.cm,file=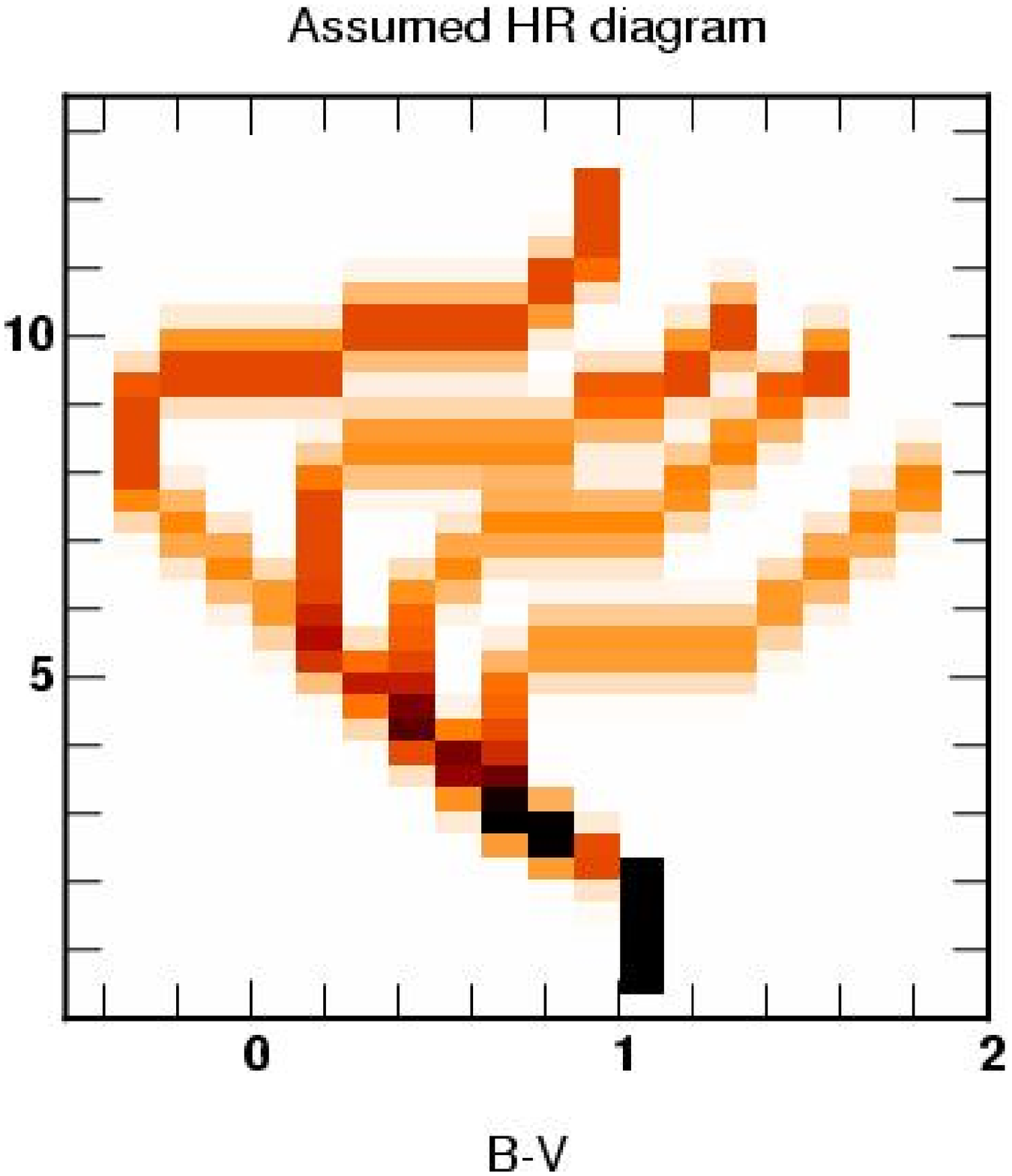}}}
\put(3.,0.5){\centerline{\psfig{width=7.cm,file=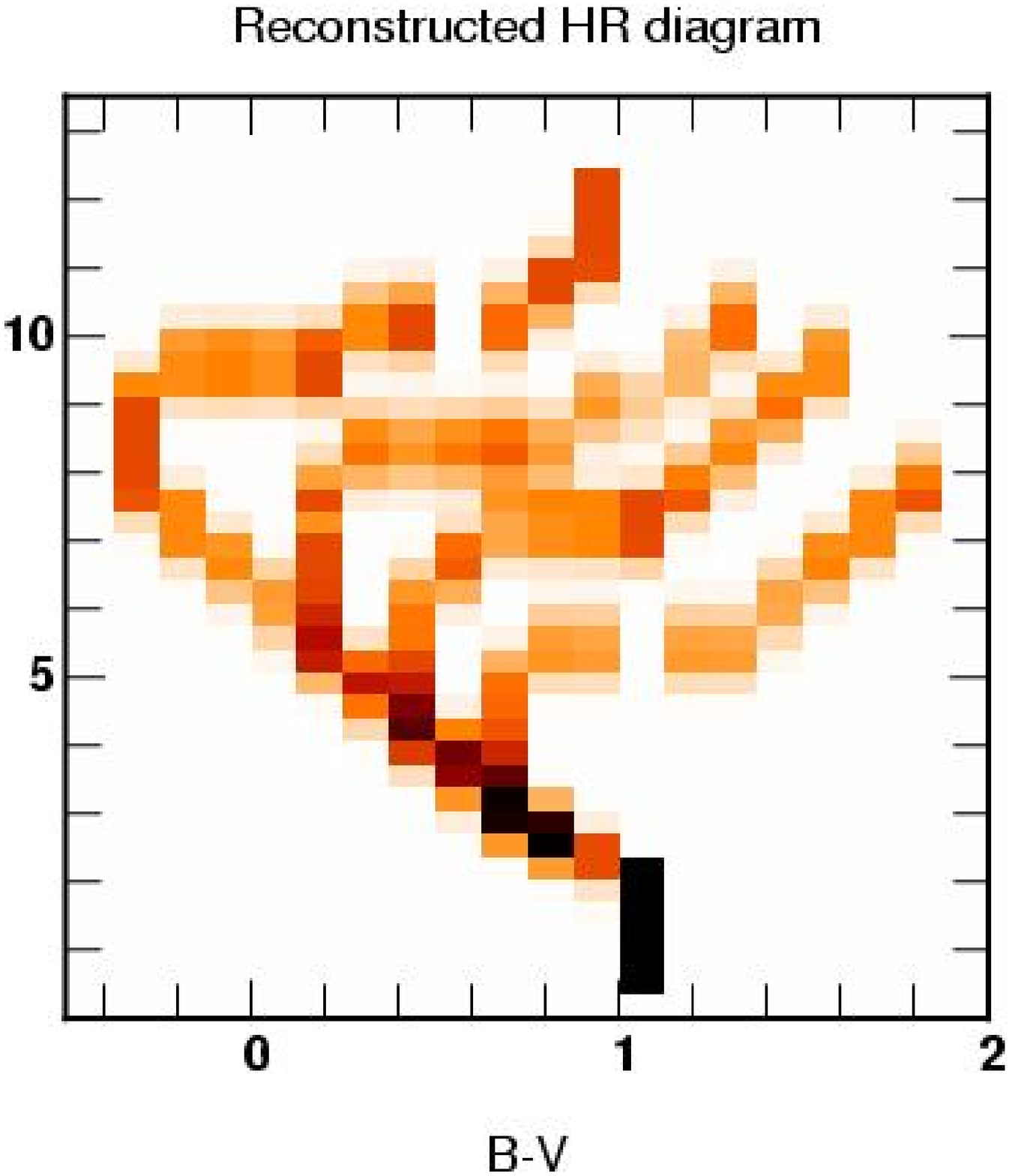}}}
\end{picture}
\caption{{\sl Left:} Assumed  HR diagram for the epicyclic  model.  The four
populations  have distinct  turnoff point  and kinematic  index.{\sl Right:}
Reconstructed HR diagram. $L_0$ is expressed in unit of $L_\odot$. Note that
all  four populations  are well  recovered. Main  sequence and  turn-off are
reconstructed within  10\% error (less  than 1\% for  the lower part  of the
main  sequence due  to the  large number  of stars  in that  part of  the HR
diagram).  Giant  branch are also recovered though  the reconstruction error
is higher.}
\label{f:model_shu}
\end{figure*}

\begin{figure*}\unitlength=1cm
\begin{picture}(14,8)
\put(-5.5,0.5){\centerline{\psfig{width=8.cm,file=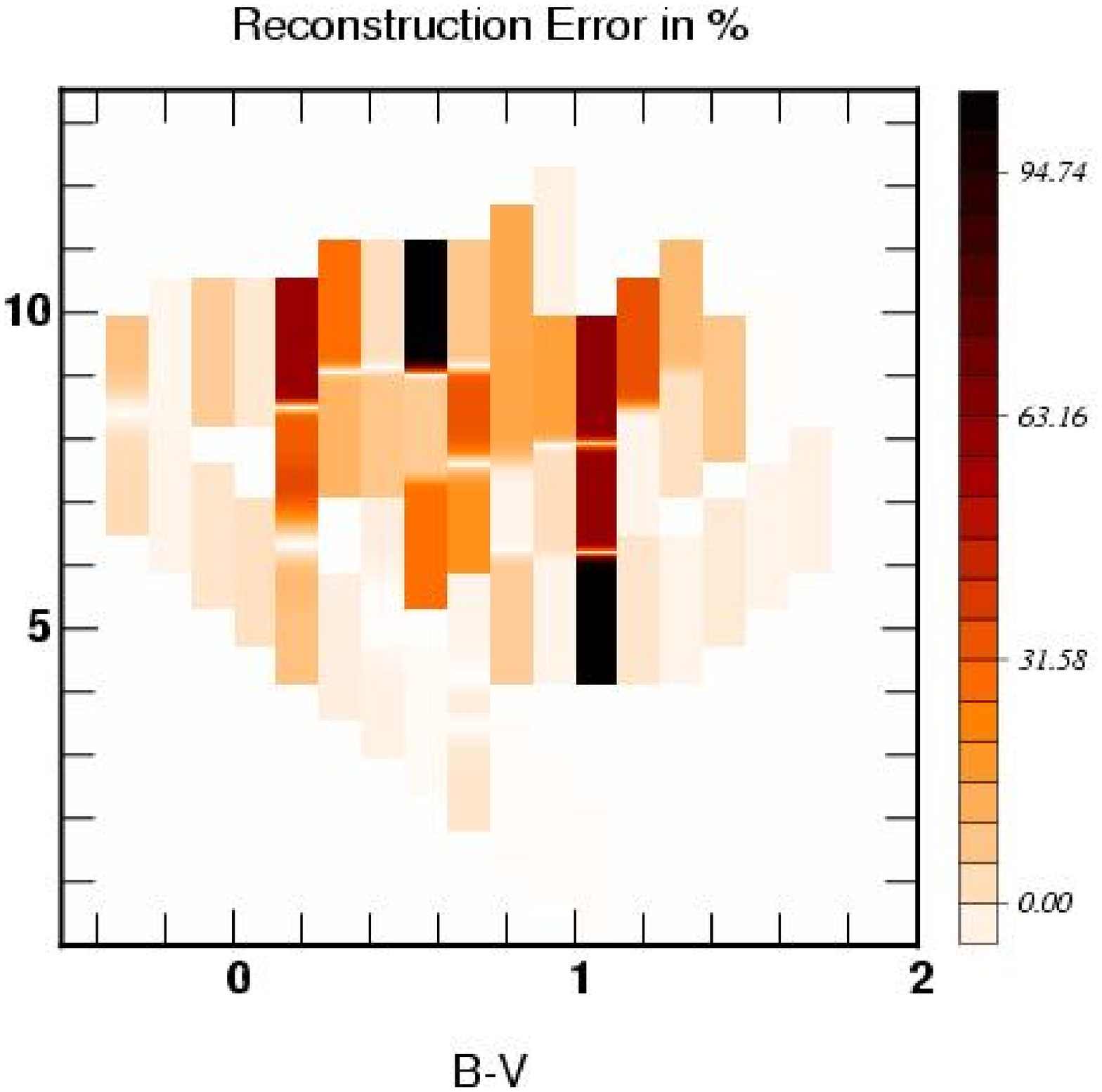}}}
\put(3.,0.5){\centerline{\psfig{width=7.cm,file=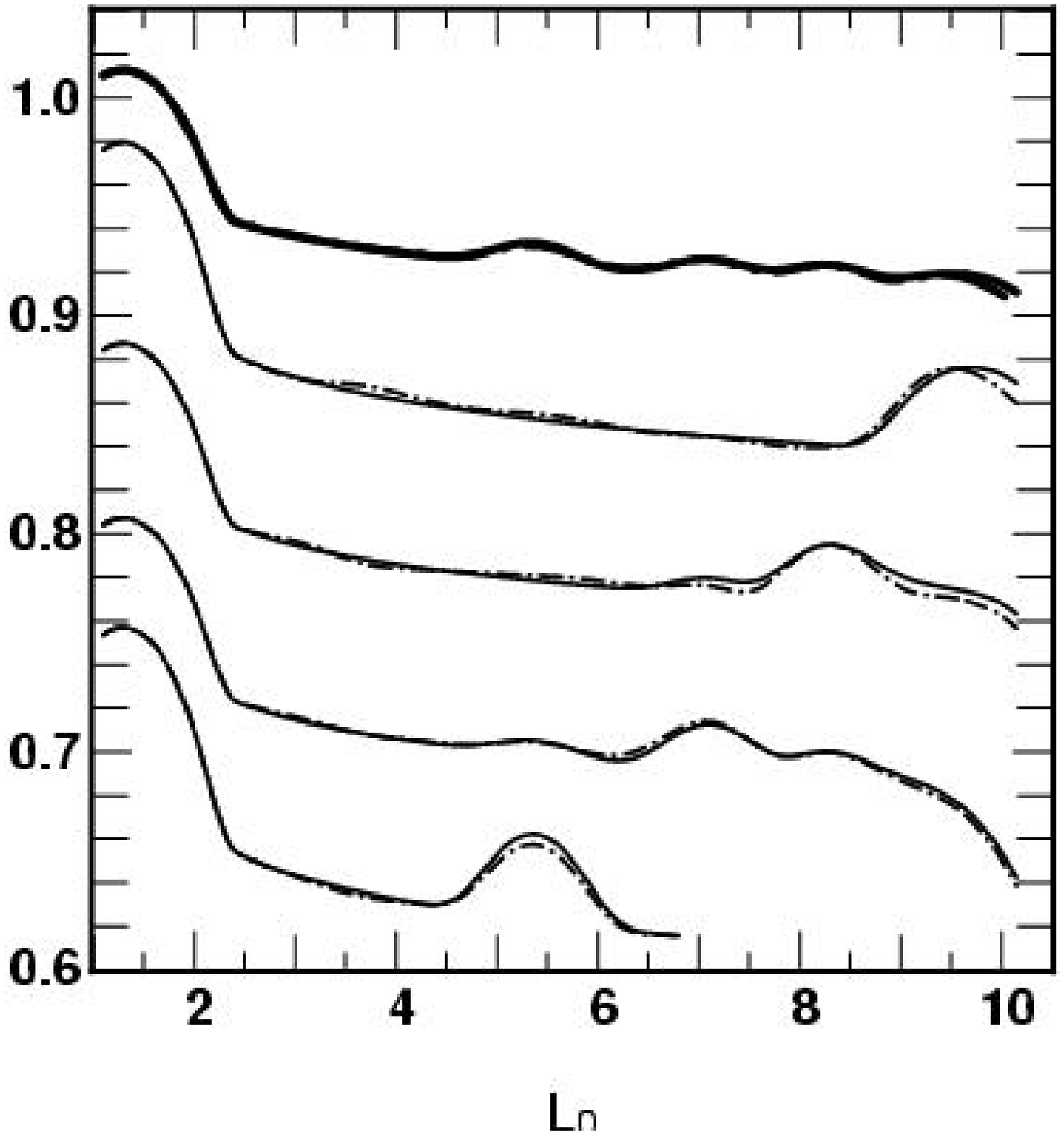}}}
\end{picture}
\caption{{\sl Left panel:} Reconstrution  error in the $[B-V,L_0]$ plane for
the epicyclic  model shown  in Figs.~\ref{f:model_shu}.  {\sl  Right panel:}
Model  vs.  recovered  luminosity  function $\Phi_\beta(L_0)$  for the  four
kinematic indexes correponding to the oldest population (lower curve) to the
youngest  (upper curve).  The  LFs are  plotted on  a logarithmic  scale and
arbitrary normalised.   The curves corresponding to the  two kinematic index
where shifted  along the  y axis.   Plain lines correspond  to the  model LF
while dash-dotted lines are the reconstructed  LF.  Note that the LF is well
reconstructed  for  the  main  sequence  (at low  luminosity)  and  for  the
turn-off. The total  luminosity function summed over the  kinematic index is
also displayed as the top thick lines.  The bumps at low and high luminosity
are properties  of the model  and correspond to  the lower part of  the main
sequence and to the subgiant branch of each population.}
\label{f:rec_shu}
\end{figure*}

{Figure~\ref{f:rec_shu}  shows the reconstruction  error in  the $[B-V,L_0]$
plane correpsonding  to the HR diagram shown  in Fig~\ref{f:model_shu}.  The
main sequence and  the different turn-off are well  reconstructed (the error
lies well below 1\% for the faint part of the main sequence and is less than
10\%  at  the  turn-offs).  The  Red  Giant  Branches  (RGB) are  also  well
reproduced even though it strongly depends on the age of the population (via
$\beta$). This can  be understood if we  look at the number of  stars in the
different regions on  the $[B-V,L_0]$ plane.  Older (  {\sl resp.}  younger)
populations have  larger ({\sl resp.}lower) number  of stars on  the RGB and
the SNR  is increasing ({\sl  resp.}  decreasing) correspondingly.   We note
that the four tracks are  recovered without creating any spurious structure.
The  luminosity   functions  $\Phi_\beta(L_0)$  are   recovered  within  1\%
uncertainty (in  mean value) for the  oldest population and  within 20\% for
the youngest (note that sometimes,  the reconstruction error increases up to
100\% when no stars are recovered on the RGB)}

\section{Discussion and conclusion}

The  main result  of  this paper  is  a demonstration  that the  generalized
stellar  statistic  equation  including  proper  motions,  \Eq{N0},  can  be
inverted, giving access to both  the kinematics and the luminosity function.
The inversion was carried  for two rather specific functional decompositions
of the underlying distribution (namely, constant ratio and possibly singular
Schwarzschild  ellipsoids  plane  parallel  models) {and  a  more  realistic
physical model (the epicyclic Shu model) which accounts for gradients}.  The
inversion  assumes  that the  departure  from  harmonicity  of the  vertical
potential,  and/or the  asymmetric  drift or  the  Sun's vertical  velocity,
$w_\odot$, are known. Indeed the break in the potential yields a scale which
reflects  the fact that  statistically the  dynamics (i.e.   the velocities)
gives  a  precise indication  of  distances in  units  of  that scale.   The
asymmetric drift or vertical component of the sun's velocity provide another
energy scale  (and therefore a distance  scale). The existence  of more than
one distance scale is  mathematically redundant, but practically of interest
for the purpose of accounting for local and remote stars.\\

In  a nutshell,  it  was  shown in  section~\ref{s:deriv}  that \Eq{N0}  has
solutions  for  families  of  distribution obeying  \Eq{ansatzPP}  (singular
ellipsoid) or \Eq{Schwarzschild}  (Schwarzschild ellipsoid). Those solutions
are  unique and  can be  made  explicit for  a number  of particular  cases:
\Eq{deconvol}  (pin-like  velocity   ellipsoid),  \Eq{SN1}  (constant  ratio
$\beta_R/\beta_{z}$,  $w_{\odot}  \approx   0$),  \Eq{N24}  (constant  ratio
$\beta_R/\beta_{z}$   and  $\beta_\phi/\beta_{z}$  either   with  $v_{\odot}
\approx {\bar  v}_{\phi}\,$,$\,w_{\odot} \approx 0  \,$, or $\,  v_{\odot} -
{\bar v}_{\phi}  \neq 0 \,$,$ \, w_{\odot}  \neq 0 \,$ \&  $\chi \approx 0$:
statistical secular  parallaxes).  In all other instances,  the solution can
be  found via the  general non-parametric  inversion procedure  described in
section~\ref{s:NP}, the  only constraint being the computation  of the model
matrix generalizing  \Eq{eqnNP} (which might  require numerical integration,
as  shown  for  instance  in  section~\ref{s:epi});  in  this  more  general
framework it  remains also to  demonstrate that the inversion  will converge
towards a  solution which is {\sl  unique}.  {For instance,  in the r\'egime
where  the epicyclic  model  has been  tested (subsection~\ref{s:simshu})  a
unique solution  seems to be well  defined. The luminosity  function of each
kinematical component is well recovered throughout the HR diagram.

  More tests are  required before
applying the  method to real  data, and are  postponed to a  companion paper
(Siebert  et al.  in  preparation).}   For a  given  vertical potential,  it
appears  that  the  modelling  of  star  counts  indexed  by  proper  motion
$A_{\lambda}(m, \mu_l,\mu_b;l,b)$ has a  solution for most model parameters.
Many different models based on distinct {\it priors} have produced realistic
magnitude  and  colour star  counts  but  failed  to predict  proper  motion
measurements accurately  (for instance, note  that the Besan\c con  model --
which relies on  a nearly dynamical consistent model --  produces a good fit
to proper motion surveys (Ojha, 1994), while dynamically inconsistent models
are more problematic (Ratnatunga, 1989)).\\

It   should  be   emphasized  that   the  inversion   method   presented  in
section~\ref{s:NP}  is a  true deconvolution  and should  give access  to a
kinematically  indexed HR  diagram. Together  with  some model  of the  time
evolution of  the different kinematic  components (via, say, a  disk heating
mechanism), the indexing could be  translated into one on cosmological time,
hence  providing a  non-parametric  measurement of  the local  neighbourhood
luminosity function which is  complementary to that obtained by evolutionary
track fitting with an assumed  initial mass function and star formation rate
(see, for  instance, Hernandez, Valls-Gabaud  \& Gilmore, 1999).  Note that,
conversely, the agreement between the  standard direct method to predict the
local luminosity  function and  the method presented  here could be  used to
measure the Galactic potential.\\

{ The deepest photometric and proper motion of whole sky survey available is
the Tycho-2 catalogue ( H\o g et  al., 2000) which is a new reduction of the
Tycho  data (H\o  g et  al., 1998).  Many Tycho  stars are  disk  giants and
subgiants covering a large range of distances; the method developed here can
be applied  to these  stars and  will allow us  to recover  their luminosity
function without  any prior information  from stellar evolution  tracks.  We
intend in  a forthcoming  paper to  apply the method  presented here  to the
Tycho-2 catalogue (H\o g et al., 1998) and to other proper motion catalogues
in  order  to  determine the  luminosity  function  of  stars in  the  Solar
neighbourhood.  We   will  investigate  the  limitations   introduced  by  a
magnitude-limited catalogue, by the  finite size of catalogues and also by
our limited knowledge of the Galactic potential.  }
{  Reddening is  also bound  to be  a concern  since it  will  bias apparent
luminosities as a  function of $\ell$ and $b$.  If  the reddening is diffuse
and the absorbing component law is  known, the kernel of {\it e.g.}  \Eq{N2}
will simply be  modified accordingly.  Alternatively, multicolour photometry
could be sufficient  to constrain the spatial extinction  law. Of course the
dimensionality of the problem is increased by the number of different colour
bands  used, since the  analysis must  be carried  while accounting  for all
colours simultaneously.  }

{ The final  error on the recovered luminosity functions  will depend on the
photometric errors of the observational catalogue($\sim$ 0.1 for the Tycho-2
catalogue  down  to  0.013  for  $V_{\rm  T}  <  9$,  $\sim  0.05-0.10$  for
photographic surveys).   It will also  depend on the relative  proper motion
accuracy   ($\delta  (m)  \sim   2  \delta   (\mu)  /   \sigma_{\mu}$,  with
$\sigma_{\mu}$  the  typical dispersion  for  a  stellar  group at  a  given
distances).  With the Tycho-2 catalogue completed by proper motions (with an
accuracy of 2.5  mas/y), and for disk giants  with velocity dispersions from
10 to 50 km/s and proper motion dispersions from 2 to 10 mas/y, the accuracy
on the recovered luminosity function will be limited to about 0.5 magnitude.
Closer (and fainter) stars  with proper motions from photographic catalogues
will  constrain the  lower part  of the  luminosity function  with  a higher
accuracy.  }

In a  decade, sky surveys by the  Fame, Diva and GAIA  satellites will probe
the Galactic  structure in superb  detail, giving directly access  to larger
volumes of  the 6D stellar  phase space of  the Galaxy. It will  remain that
farther  out,  only  proper  motions  and photometry  will  have  sufficient
accuracy and  generalization of  methods such as  that derived here  will be
used to extrapolate  our knowledge of the kinematic  and luminosity function
of  the Galaxy  and its  satellites. For  instance Appendix~\ref{s:external}
sketches the  possible inversion of an external  globular cluster luminosity
function with GAIA quality photometry.

\section*{Acknowledgements}
{\em We  would like  to thank J.L.~Vergely for early stimulating 
discussions on this project and E.~Thi\'ebaut for  fruitful comments. 
}

\def\BIBITEM#1#2#3{\bibitem[\protect\citename{#2, }#3]{#1}}

\def\aj {Astronomical Journal}
\def\apj{ApJ }
\def\apjs{ApJS }
\def\pasj{Publ. Astron. Soc. Japan}
\def\aj{AJ }
\def\mn{MNRAS }
\def\pasp{PASP }
\def\apjl{ApJL \rm}
\def\mnras{Mon. Not. R. Astron. Soc}
\def\aeta{A\&A }
\def\aetal{A\&AL }


\appendix

\section{Asymptotic analytic solutions for the Schwarzschild model}
\label{s:slices}

Let us demonstrate that \Eq{N0} has explicit analytic solutions for families
of  distribution obeying \Ep{Schwarzschild},  using the  inversion procedure
sketched in section~\ref{s:toy}.

\subsection{Slices towards the Galactic centre}

For  the sake  of simplicity  let us first
restrict the  analysis  to $u_{\odot}=
v_{\odot} =w_{\odot}=0 $  and assume first we have  measurements only in the
direction $\ell=0$.  The integration over $u_{r}$ then yields
\begin{equation}
\int     f_\beta(\M{r},\M{u})   \d u_{r}=  \sqrt{\frac{\beta_{b}\beta_{\phi}
   }{4 \pi^{2}}}      
   \exp\left(-\frac{1}{2}\left[ \beta_{\phi} (u_{\ell}-{\bar v_{\phi})}^{2}
   +
 \beta_{b}
   u_{b}^{2} \right]  {-\beta_{z}      \psi_{z}(z)  }  \right)
 \,,\EQN{Schwarzschild2}
\end{equation}
where 
\begin{equation}
\beta_b^{-1}   =\beta_R^{-1}    \sin^2(b)    +      \left(\frac{r_\odot   -r
\cos(b)}{R}\right)^2 \beta_z^{-1} \cos^2(b) \,. \EQN{beta-1}
\end{equation}

Without loss of generality let us integrate over $u_{\ell}$
\begin{equation}
\int  \!\!  \int  f_\beta(\M{r},\M{u})       \d  u_{r}  \d      u_{\ell}   =
 \sqrt{\frac{\beta_{b}}{2\pi^{}}}   
  \exp\left(-\frac{1}{2}{ \beta_{b} u_{b}^{2}} {-\beta_{z} 
  \psi_{z}(z)  }\right) \, .\EQN{Schwarzschild3}
\end{equation}
At large  distances to  the Galactic centre,  both  $R$ and  $r_{\odot}$ are
large compared to $r$ and \Eq{beta-1} becomes
\begin{equation}
{\bar \beta}_b^{-1} =\beta_R^{-1} \sin^2(b) + \beta_z^{-1} \cos^2(b) \,.
\EQN{barbeta}
\end{equation}

Let us now also assume  that $\beta_{R}$ and $\beta_{z}$ are known monotonic
functions  of   a   unique   parameter   $\beta$.  We   may   now   convolve
\Eq{Schwarzschild3} with  the luminosity function sought,  $\Phi\left[ L r^2
,\beta \right]$ so that
\begin{equation}
A[b,\mu_b ,L]= {\int \!\! {\int { 
\sqrt{ {\frac{{\bar \beta}_{b}}{2\pi }}}
\Phi \left[ L r^2 ,\beta    \right]}}\exp \left(  
    {-  \frac{1}{2} r^{2} {\bar \beta}_{b} \mu_b^{2}
    -\beta_{z}  \psi_{z}(
r \sin(b))}  \right)r^3   \d r\d\beta  }  \,.
\EQN{SN1}
\end{equation}

\Eq{N1} appears now as a special case of \Eqs{barbeta}{SN1} corresponding to
$\beta_{R} \rightarrow \infty$.  Even though the  convolution in \Eq{SN1} is
less straightforward than that of \Eq{N22}, and so long as $\psi_{z}$ is not
purely harmonic, \Eq{SN1}  will have a non-trivial  solution for $\Phi$.  In
particular, if the  ratio of  velocity dispersions $\beta_{R}/\beta_{z}$  is
assumed  constant, \Eq{N22} still holds  but with $\beta=\beta_{z}$, and $x$
replaced by $x'$ defined by
\[
x' = 
  \alpha \frac{\sin^2 (b)}{L}+ {{\mu_b ^2} \over { 2 L \cos^2 (b) + \xi 2 
  L  
  \sin^2 (b)}} \,,\quad {\rm where} \quad \xi = 
  \frac{\beta_{z}}{\beta_{R}}\, ,
\quad \mbox{
with} \quad
A'[b,\mu_{b},L] = A[b,\mu_{b},L] \sqrt{1+\xi \tan^{2}(b)} \,.
\]
Note that if $v_\odot$ and $w_\odot$ are not negligible, \Eq{SN1} becomes
\begin{equation}
A[b,\mu_b ,L]= {\int \!\! {\int  { \sqrt{ {\frac{{\bar \beta}_{b}}{2\pi  }}}
\Phi   \left[  L r^2   ,\beta    \right]}}\exp \left( {-  \frac{1}{2}  {\bar
\beta}_{b} (u_b+ \cos(b) w_\odot)^{2}    -\beta_{z} \psi_{z}( r    \sin(b))}
\right)r^3 \d r\d\beta } \,, \EQN{SN11}
\end{equation}
and is of the form discussed below as \Eq{N24} with $\ell =0$.

\subsection{Slices away from the Galactic centre}

For any direction  $\ell\ne0$ when
$r_\odot \!\rightarrow \!
\infty$, the kinetic  dispersion (replacing in \Eq{barbeta}) along Galactic
latitude is then given by : 
\[
 {\hat \beta}_b^{-1} =\left( \beta_{R}^{-1} \cos^2 \ell + \beta_{\Phi}^{-1}
\sin^2 \ell \right) \sin^2 b + \beta_{z}^{-1} \cos^2 b  \,,
\]
and \Eq{N1} is replaced by
\begin{equation}
A[b,\ell,\mu_b  ,L]= {\int \!\!   {\int { \sqrt{ {\frac{\hat \beta_{b}}{2\pi
}}}  \Phi \left[ L r^2  ,\beta \right]}}\exp \left(  {-  {\hat \beta}_{b} (r
{\mu_b}+ \cos(b)  w_\odot-\sin(b)\ \sin(l)\ [{v_\odot}-{\bar v}_{\phi}])^{2}
- -\beta_{z} \psi_{z}( r \sin(b))} \right)r^3 \d r\d\beta } \,, \EQN{SN2}
\end{equation}
which can be rearranged as (again with $\beta=\beta_{z}$)
\begin{equation}
L^2     \cos(b)     A_{2}[b,\ell,\mu_b     ,L]=      {\int    \!\!     {\int
 \sqrt{\frac{\beta}{2\pi}} {\Phi\left[ u^2,\beta \right]}}\exp \left( -\beta
 u^2  x_{2} +\beta u y_{2} -  \beta z_{2} -  \beta \chi(u y) \right)u^3 \d u
 \d\beta } \,, \EQN{N24}
\end{equation}
with $z_2$ given by \Eq{defz2}
\begin{equation}
x_{2} = \alpha \frac{\sin^2 (b)}{L}+ {{\mu_b ^2} \over {2 L \cos^{2}(b)+ 2 L
  \sin^{2}(b)\left( \xi_{R}\cos^{2}(\ell)+ \xi_{\phi}\sin^{2}(\ell)\right)}}
  \,,
\end{equation}
\begin{equation}
y_{2} = {{ \mu_{b} (  [{v_\odot}-{\bar v}_{\phi}] \sin  b \sin \ell - w_\odot
 \cos b)} \over    {  \sqrt{  L} \cos^{2}(b)+   \sqrt{L}   \sin^{2}(b)\left(
 \xi_{R}\cos^{2}(\ell)+ \xi_{\phi}\sin^{2}(\ell)\right)}} \,,
\end{equation}
\begin{equation}
A_{2}[b,\ell,\mu_{b},L]   =  A[b,\mu_{b},L]     \sqrt{1+        \tan^{2}(b)[
\xi_{R}\cos^{2}(\ell)+  \xi_{\phi}\sin^{2}(\ell)]}  \,, \quad  {\rm   where}
\quad  \xi_{R}  =    \frac{\beta_{z}}{\beta_{R}}\,  ,   \quad   \xi_{\phi} =
\frac{\beta_{z}}{\beta_{\phi}}\, .
\end{equation}

In the region where  the asymmetric drift and the  z-component of  the Sun's
velocity can be neglected, ${v_\odot} \approx  {\bar v}_{\phi}$ and $w_\odot
\approx 0$, $y_{2}$ and $z_{2}$ vanish and \Eq{N24} is formally identical to
\Eq{N2}; once again the solution of \Eq{N24}  is given by \Eq{deconvol} with
the appropriate substitutions.   Alternatively, in the regions  where either
$w_\odot$ or ${v_\odot}-{\bar v}_{\phi}$ cannot be neglected, \Eq{N24} has a
unique solution  even if $\chi   \equiv 0$,  which can  be  found along  the
section $\mu_b =0$  (Note  that when  $r_\odot \rightarrow \infty$,  we  can
always assume  $u_\odot =0$ by  changing  the origin  of Galactic longitude,
$\ell$). Indeed, \Eq{N24} becomes \Eq{N25mt}
%
which is of the form described  in section~\ref{s:unique}  with  $\nu=0$,
$x_{3}$ replacing $x$ and $z_{2}$ replacing  $y$; the corresponding solution
is found following  the same route.   It is analogous to statistical secular
parallaxes (note nonetheless that the section $\mu_{b}=0$ might not 
be sufficient to carry the inversion without any truncation bias since 
$\log(z_{2})$ spans 
$]-\infty,Z[$ when $b$ and $\ell$ vary with $Z$ a function of $\xi_{R}$,$\xi_{\Phi}$,
$w_\odot$ and ${v_\odot}-{\bar v}_{\phi}$). 

Turning back to \Eq{N24},  it remains  that for more  general $\chi$  it can
still be inverted via  the  kernel, $K_{2}(x_{2},y_{2},z_{2},y |  u,\beta)$,
which depends explicitly on $\chi$:
\[
K_{2}(x_{2},y_{2},z_{2},y |  u,\beta)=  \sqrt{\frac{\beta}{2\pi}} 
\exp\left( -\beta  u^2 x_{2}     +\beta u y_{2} - \beta z_{2} 
- \beta \chi(u y)  \right)u^3 \, .
\]
Note that the multi-dimensionality of the kernel, $K_{2}$, is not a 
problem from the point of view of a $\chi^{2}$ non-parametric 
minimisation described in section~\ref{s:simul}.

\section{External  spherical isotropic clusters}
\label{s:external}

Consider a  satellite of our   Galaxy  assumed to be    well-described as  a
spherical  isotropic   cluster with a luminosity    function indexed by this
kinematic temperature.
Let  $4 \pi^2 A_\lambda(\mu_b ,L,R) \mu d \mu \d L R \d R$  be the number of
stars which have proper   motions,  $\mu^2= \mu_b^2+\mu_\ell^2$ ,   apparent
luminosity $L$ at radius $R$ from the centre at the wavelength $\lambda$.
This quantity   is      a  convolution   of the      distribution   function
 $f(\varepsilon,\beta)$ (a   function of energy,  $\varepsilon$,  and $\beta
 \equiv 1/\sigma^2$) and the  luminosity function, $g_\lambda(L_0,\beta)$, a
 function of  the intrinsic luminosity, $L_0$,  the population, $\beta$, and
 wavelength $\lambda$:
\begin{equation}
A_\lambda(\mu,  L ,R) =  \int     \!\! \int \!\!\int    f(\varepsilon,\beta)
g_\lambda(\beta, L r'^2)\, \d \beta \, { \d z}\, { \d v_{z}} \,,
\end{equation}
which can be rearranged as
\begin{equation}
A_\lambda(\mu,  L ,R)   =  4  \int  \!\! \int  \!\!\int   f(\varepsilon,\beta)
g_\lambda(\beta, L   r'^2)    \frac{r   \d  r}{\sqrt{r^2-R^2}}   \frac{   \d
\varepsilon}{\sqrt{2(\psi + \varepsilon)-v^2 }} \d \beta   \,,
\EQN{sphericalStarCount}
\end{equation}
where $v^2= \mu^2 r'^2$ is the velocity in the  plane of the sky, and $r'^2=
(r^2-R^2  +   r_\odot^2)$ the  distance to  the   observer and $r_\odot$ the
distance  to the cluster, and  $r$ the distance to the  cluster centre.  The
potential  can   be derived non-parametrically  from   the projected density
(using Jeans' equation). Indeed the enclosed mass within  a sphere of radius
$r$ reads
\begin{equation}
M_{\rm dyn}(<r)  =  r^{2} \frac{\d \psi}{\d r   } = -\frac{r^{2}}{\rho} \frac{\d
{(\rho \sigma^2) }}{\d r } \,, \EQN{mass0}
\end{equation}
where $\psi(r)$ is the gravitational potential $\rho(r)$ the density and 
$\sigma(r)$ the radial velocity dispersion. The surface density is related to 
the density via an Abel transform:
\begin{equation}
\Sigma(R) = \int\limits_{-\infty}^{\infty} \rho(r)  \d z = 2 \int_{R}^\infty
 \rho(r) \frac{r   \d  r }{\sqrt{r^2-R^{2}}}  \equiv {\cal  A}_{R}(\rho)\, ,
 \EQN{Sigma}
\end{equation}
where $\Sigma(R)$ is the projected surface density and $R$ the projected
radius as measured on the sky. Similarly the projected velocity
dispersion $\sigma_{p}^{2}$ is related to the intrinsic velocity
dispersion, $\sigma^{2}(r)$, via the \textsl{same} Abel transform 
(or projection)
\begin{equation}
\Sigma(R) \sigma_{p}^{2}(R)=2 \int_{R}^\infty \rho(r) \sigma^{2}(r)
\frac{r \d r }{\sqrt{r^2-R^{2}}} \equiv {\cal A}_{R}(\rho  \sigma^{2}) \, . 
\EQN{Sigmau2}
\EQN{epsilon}
\end{equation}
Note that $\Sigma(R) \sigma_{p}^{2} $ is the projected kinetic  energy
density divided by three (corresponding to one degree of freedom) and
$\rho(r) \sigma^{2}$ the kinetic energy density divided by three.
Inserting \Eqs{Sigma}{Sigmau2} into \Eq{mass0} yields:
\begin{equation}
M_{\rm dyn}(<r)    = -\frac{r^{2}}{{\cal  A}^{-1}_{r}(\Sigma)} \frac{\d
{{\cal A}^{-1}_{r}(\Sigma \sigma_{p}^{2})}}{\d r } \, ,\quad {\rm while}\quad
\rho(r)= \frac{1}{4 \pi r^2}\frac{\d {}}{\d r} M_{\rm dyn}(<r)
\quad {\rm and}
\quad  \nabla^{2} \psi = -4 \pi G \rho
 \, . \EQN{mass}
\end{equation}  
The underlying isotropic distribution is given by
an inverse Abel from the density.   
\begin{equation}
f(\varepsilon) =\frac{1}{\sqrt{8}\pi^2}\int \frac{\d{}^2 \rho}{\d
\psi^2}\frac{\d \psi}{\sqrt{\varepsilon-\psi}} 
\equiv \int\limits_0^\infty F(\beta)
\exp(-\beta \varepsilon) \d \beta \,
\end{equation}
where  an isothermal decomposition over  temperature $\beta$ was assumed for
the distribution function (this assumption is not required: any parametrized
decomposition is acceptable). So
\begin{equation}
F(\beta) = {\cal L}^{-1} \left[f(\varepsilon)\right] =
{\cal L}^{-1} \left[{\cal A}^{-1} \left({\cal A}^{-1} \left( \Sigma\right)\right)\right]
\end{equation}
where $\cal L$ is the Laplace operator.

Calling
\begin{equation}
G[Y]=\int\limits_0^Y
 \frac{\exp(-X) \, \d x}{\sqrt{Y- X }}={{{\sqrt{\pi }}\,
     \rm{Erfi}({
         \sqrt{Y}})}{{e^{-Y}}}}\,, \quad   g_1(\beta, L
 r'^2)=
g_\lambda(\beta, L r'^2) F(\beta) \beta^{-3/2}
\end{equation}
\Eq{sphericalStarCount} becomes 
\begin{equation}
A_\lambda(\mu,  L   ,R)    =  2    \sqrt{2}   \int\limits_0^\infty  \left(
\!\!\int_R^\infty   G\left[\beta(\frac{\mu^2(r^2-R^2     +    r_\odot^2)}{2}
- -\psi(r))\right] g_1\left[\beta,  L (r^2-R^2 + r_\odot^2)\right]  \frac{r \d
r}{\sqrt{r^2-R^2}} \right) \d \beta \EQN{external-final}
\end{equation}
where $G$ is a known kernel while $g_\lambda$ is the unknown sought
luminosity function. \Eq{external-final} is the direct analogue to
\Eq{N22}. It will be invertible following the same route with GAIA
photometry.
(With today's accuracy in photometry, for a typical globular cluster at a
distance, $r_{\odot}$ of, say $10$ kpc, the relative positions within the
cluster are negligible w.r.t $r_{\odot}$:
\(
r^2-R^2 \ll r_\odot^2 \,,
\)
therefore 
\[
A_\lambda(\mu,  L   ,R)    =  2   \sqrt{2}   \int\limits_0^\infty   \left(
\!\!\int_R^\infty  G\left[\beta(\frac{\mu^2 r_\odot^2 }{2}  -\psi(r))\right]
\frac{r \d  r}{\sqrt{r^2-R^2}} \right) g_1\left[\beta, L r_\odot^2 \right]\d
\beta 
\]
$L$ is then  also mute, and the inversion problem shrinks to one involving
finding the relative weights, $g_L\left[\beta\right]$ of a known
distribution).

\end{document}